\documentclass[3p,times]{revtex4}%{elsarticle}
\usepackage{amsmath,float,amssymb,amsbsy,gensymb,amstext,amsthm,booktabs,simplewick,exscale,relsize,graphicx,amsfonts,upgreek,slashed,color,multirow,dcolumn,bm,enumerate,url,hyperref}

\def\ep{\epsilon}

\begin{document}
%%%%%%%%%%%%%%%%%%%%%%%%%%%%%%%%%%%%%%%%%%%%%%%%%%%%%%%%%%%%%%%%%%%%%%%%%%%%%%
\title{\mbox{}\\[10pt]
 Probing Lorentz violation effects via a laser beam interacting with a high-energy charged lepton beam}

%%%%%%%%%%%%%%%%%%%%%%%%%%%%%%%%%%%%%%%%%%%%%%%%%%%%%%%%%%%%%%%%%%%%%%%%%%%%%%
\author{Seddigheh Tizchang~\footnote{s.tizchang@ipm.ir}}
 \address{School of Particles and Accelerators, Institute for Research in Fundamental Sciences (IPM) P.O. Box 19395-5531, Tehran, Iran\vspace{0.2cm}}

\author{Rohoollah Mohammadi~\footnote{rmohammadi@ipm.ir}}
\address{Iranian National Museum Of Science and Technology (INMOST) P.O. Box 11369-14611, Tehran, Iran\\
	School of Astronomy, Institute for Research in Fundamental Sciences (IPM)
	P. O. Box 19395-5531, Tehran, Iran\vspace{0.2cm}}

\author{ She-Sheng Xue~\footnote{ xue@icra.it}}
\address{ ICRANet and Department of Physics, University of Rome``Sapienza''  P.le A. Moro 5, 00185 Rome, Italy\vspace{0.2cm}}

%\date{\today}

%%%%%%%%%%%%%%%%%%%%%%%%%%%%%%%%%%%%%%%%%%%%%%%%%%%%%%%%%%%%%%%%%%%%%%%%%%%%%%
\begin{abstract}
    In this work, the conversion of linear polarization of a laser beam to circular one  through its forward scattering by a TeV order charged lepton beam in the presence of Lorentz violation correction is explored. We calculate the ratio of circular polarization to linear one (Faraday Conversion phase $\Delta\phi_{\rm{FC}}$) of the laser beam interacting with either electron or the  muon beam in the framework of the quantum Boltzmann equation. %^We  discuss how the interaction of laser beam and charged lepton beams can provide a suitable situation to constrain Lorentz violation coefficients, especially for $c_{\mu \nu}$ coefficients.
     Regarding the experimentally available sensitivity to the Faraday conversion $\Delta\phi_{\rm{FC}}\simeq 10^{-3}-10^{-2}$, we show that the scattering of a linearly polarized laser beam with energy $k_0\sim 0.1$ eV  and an electron/muon beam with flux $\bar{\epsilon}_{e,\mu}\sim 10^{10}/10^{12}$ TeV cm$^{-2}$ s$^{-1}$ places an upper bound on the combination of lepton sector Lorentz violation  coefficients $c_{\mu\nu}$ components $(c_{TT}+1.4~c_{(TZ)}+0.25(c_{XX}+c_{YY}+2~c_{ZZ})$. The obtained bound on the combination for the electron beam  is at the $4.35\times 10^{-15}$ level and for the muon beam  at the $3.9\times 10^{-13}$ level. It should be mentioned that the laser and charged lepton beams considered here to reach the experimentally measurable $\Delta\phi_{\rm{FC}}$ are currently available or will be accessible in the near future.  This study provides a valuable supplementary to other theoretical and experimental frameworks for measuring and constraining Lorentz violation coefficients.
\end{abstract}

%\begin{keyword} 
%\end{keyword}
%%%%%%%%%%%%%%%%%%%%%%%%%%%%%%%%%%%%%%%%%%%%%%%%%%%%%%%%%%%%%%%%%%
\maketitle
\section{ Introduction}\label{sec:I}
Usually, radiation can be both linearly and circularly polarized.  It is well known that when initially unpolarized photon scatters off a free electron through Compton scattering, it results in linear polarization but not circular polarization of the scattered radiation. However, it is shown that Compton scattering in the presence of external background fields similar to strong magnetic fields \cite{{Cooray:2002nm},{Bavarsad:2009hm},{Motie:2011az},{Shakeri:2017iph}} or theoretical (non-trivial) backgrounds such as non-commutativity in space-time   
\cite{{Bavarsad:2009hm},{Tizchang:2016hml},{Batebi:2016ocb}} and Lorentz symmetry violation \cite{Bavarsad:2009hm}, can produce  circular polarization. Moreover, nonlinear effects such as the nonlinear Euler-Heisenberg effect can cause converting of photons from linear polarization to circular polarization \cite{{Motie:2011az},{Mohammadi:2014nca}, {Sadegh:2017rnr}}. In this paper, we consider Compton scattering through the collision of laser photons and high-energy charged lepton beams in the presence of Lorentz violation (LV) effects to study the generation of circular polarization in an Earth-based laboratory.
Lorentz symmetry is a fundamental symmetry of the standard model in flat space-time and quantum field theory. However, it can be violated by an underlying theory at Planck scale \cite{{Kostelecky:1989jp},{Bluhm:2004ep}}. There are many theories in which the Lorentz symmetry is violated spontaneously, such as string theory \cite{{Kostelecky:1988zi},{Mavromatos:2007xe},{Hashimoto:2014aoa}}, quantum gravity \cite{{Gambini:2011nx},{Moffat:2009ks},{Mavromatos:2010ar},{Collins:2006bw}} and non-commutative space-time \cite{{Calmet:2004ii},{Luo:2016bwp},{Carroll:2001ws}}.  Meanwhile, it is also possible to study LV in  a general  model-independent  way in the context of effective field theory  known as  the Standard Model extension (SME). In  the SME Lagrangian, the observer Lorentz symmetry (i.e. change of coordinate) is obeyed while the particle Lorentz symmetry (i.e. boosts on particles and not on background fields) is violated \cite{{Colladay:1996iz},{Tasson:2014dfa}}. The SME contains all feasible Lorentz breaking operators created by known fields of the standard model of dimension three or more. These operators are contracted with coefficients representing backgrounds and preferred directions in space-time \cite{{Kostelecky:2003fs},{Colladay:1996iz},{Colladay:1998fq}} and can describe small Lorentz symmetry violation at available energies.
 Generally, the number of coefficients is infinite. By the way, it is possible to  choose a minimal subset of the SME with finite coefficients. The minimal SME  contains renormalizable operators which are invariant under  the gauge group of  the  standard model, $SU(3)\times SU(2)\times U(1)$. In recent years,  new studies have provided new types of constraints on the LV parameters \cite{{Aghababaei:2017gdd},{Aghababaei:2017bei},{Santos:2018siy},{Lambiase:2017adh},{Ding:2016lwt}}. Among them astrophysical \cite{{Konnig:2016idp}, {Wei:2016ygk},{Stecker:2014oxa}} and Earth \cite{{Aaij:2016mos},{Araujo:2015zsa}, {Abe:2015ibe}} systems have shown stronger bounds on the LV parameters \cite{Kostelecky:2008ts}. 
Observation of circularly polarized photons in lepton and photon scattering can be a proof of LV and might result in new physics beyond the standard model.  In contrast, constraints on circular polarization might improve the available bounds on the parameters of the SME. \par
The paper is organized as follows: In section \ref{sec:II} we briefly introduce the Stokes parameter formalism and the generalized Boltzmann equation. In section \ref{sec:III} we study the effect of LV on the collision of the relativistic lepton beam (electron/muon) and the laser. In section \ref{sec:IV}  we give the value of the FC phase of the laser beam through this interaction. Finally, we discuss the results  in the last section.
%%%%%%%%%%%%%%%%%%%%%%%%
\section{Stokes parameters and quantum Boltzmann equation}\label{sec:II}
%%%%%%%%%%%%%%%%%%%%%%%%
Normally, the polarization of a laser beam can be described by the  well-known Stokes parameters coming in four dimensions, I, Q, U and V.  I denotes the intensity of the laser beam,  V shows the difference between left- and right- circular polarization  Q and U indicate the linear polarization. Q is defined  by the intensity difference between the polarized components of the electromagnetic wave in the direction of the $x$ and $y$ axes.  U quantifies the discrepancy between $45\degree$ and $ 135\degree$ counted from the positive $ x$  axis, to the reference plane  \cite{Kosowsky:1994cy}. The linear polarization can also be shown by  vector $P\equiv \sqrt{Q^2+U^2}$ \cite{jackson}. The Stokes parameters can be
specified  by a superposition of two opposite,  right- and left-hand,  circular polarization contributions, $(\hat{\bf R})$ and  $(\hat{\bf L})$,:
\begin{eqnarray}\label{E-componentLR}
E_L =E_{0L} \cos[\omega_0 t-\phi_L],\,\,E_R =E_{0R} \cos[\omega_0 t-\phi_R].
\end{eqnarray}
The Stokes parameters can be defined as
\begin{eqnarray}\label{eq:ST parameter}
I&\equiv&  \langle E_L^2\rangle+\langle E_R^2\rangle, \nonumber \\
Q &\equiv& \langle 2E_LE_R \cos (\phi_R-\phi_L) \rangle \, , \nonumber \\
U &\equiv& \langle 2E_LE_R \sin (\phi_R-\phi_L)  \rangle \, ,\nonumber \\
V&\equiv&  \langle E_R^2\rangle-\langle E_L^2\rangle,
\label{eqn:circstokes}
\end{eqnarray}
where the phase shift ($\Delta\phi=\phi_R-\phi_L$) is phase difference  between left- and right- hand circularly polarized waves.
This phase shift is known as a Faraday rotation (FR); it causes mixing of the Stokes parameters Q and U as follows:
\begin{equation}\label{eq:FR}
\dot U=-2Q\frac{d\Delta\phi_{FR}}{dt}\,\,\,\,\,\,\,\,\,\,\,\,\,\,\text{and}\,\,\,\,\,\,\,\,\,\,\,\,\,\,\dot Q=-2U\frac{d\Delta\phi_{FR}}{dt},
\end{equation}
 If the linearly polarized light propagates via the cold magnetized plasma, its polarization rotates and Faraday rotation will occur. Simultaneously, charged particles passing through the cold plasma emit cyclotron radiation which is circularly polarized  \cite{Kosowsky:1994cy}.
Now the phase difference($\Delta\phi=\phi_R-\phi_L$) results in mixing between the U and V Stokes parameters, known as Faraday conversion(FC).
The evolution of the Stokes parameter V given by this mechanism is obtained \cite{{Cooray:2002nm},{Ruszkowski:2001pt}}:
\begin{equation}\label{eq:FC}
\dot{V}=2\:U\frac{d\Delta\phi_{FC}}{dt},
\end{equation}
where $\Delta\phi_{FC}$ is the FC phase. Generally,  light traversing a relativistic plasma undergoes both  FC and FR.\par
Let us consider an ensemble of photons which is described by the density matrix $\rho_{ij}\equiv (|\epsilon_i><\epsilon_j|/tr\rho)$.
The density matrix can be written based on the Stokes parameters as
\begin{eqnarray}\label {eq:density matrix}
\rho=\frac{1}{2}\left(
\begin{array}{cccc}
I+Q & U-i V \\
U+iV & I-Q  \\
\end{array}
\right).
\end{eqnarray}
The density matrix $\rho_{ij}$ is related to the number operator $\mathcal{\hat D}_{ij}({\bf p}) = \hat a_i^\dagger({\bf p})\hat a_j ({\bf p})$ as
\begin{equation} \label{eq:3.17}
\langle \mathcal{\hat D}_{ij}({\bf p}) \rangle = (2\pi )^3 2p^0\delta^{(3)}(0)\rho_{ij}({\bf p}).
\end{equation}
In order to figure out the time evolution of the density matrix (Stokes parameters), it is convenient to  use the Heisenberg equation
\begin{eqnarray}
\frac{d}{dt}\mathcal{D}^0_{ij}({\bf p})=i[H_I,\mathcal{D}^0_{ij}({\bf p})],
\label{Hi}
\end{eqnarray}
where $H_I$ is the interacting Hamiltonian of photons with the standard model particles.
Obtaining the expectation value of  Eq. (\ref{Hi}) and replacing with Eq. (\ref{eq:3.17}),  the Boltzmann equation for the number operator of photons in terms of the density matrix elements $\rho_{ij}$  and  Stokes parameters  can be written as \cite{Kosowsky:1994cy}
\begin{eqnarray}
\label{eq:density-H}
\hspace{-0.8cm}
(2\pi )^3 \delta^3 (0)2\,p^0 \frac{d} {dt} \rho_{ij}(0,{\bf p})= i\langle [H_I^0(t) , \mathcal{\hat D}_{ij}({\bf p})] \rangle
-\int dt \langle [H_I^0(t),[H_I(0) , \mathcal{\hat D}_{ij}({\bf p})]]\rangle .
\end{eqnarray}
Here we consider only the contribution of photons and charged leptons to the interacting Hamiltonian.
The first term on the right-handed
side is a forward scattering term  and the second one is a higher order collision term.
Forward scattering means most of the photons travel straightforwardly without changing the momentum. Considering Compton scattering as the dominant process in the standard model, the time evolution of the circular polarization is zero, $\dot {V}=0$. However, we will show in the following that the SME as the background for Compton scattering can generate circular polarization.
In general, the interacting Hamiltonian for the photon-charged lepton beam  interaction up to leading order is given by \cite{Kostelecky:2003fs}
\begin{eqnarray}\label{eq:interactingH}
H^0_I = \int d\mathbf{q} d\mathbf{q'} d\mathbf{p} d\mathbf{p'} (2\pi)^3\delta^3(\mathbf{q'} +\mathbf{p'} -\mathbf{p} -\mathbf{q} ) \exp[i(q'^0+p'^0-q^0-p^0)t]\left[b^\dagger_{r'}a^{\dagger}_{s'}(\mathcal{M})a_sb_r\right],\label{h0}
\label{H}
\end{eqnarray}
where  $b_r (\textbf{p})$  and $b_{r'}^\dagger(\textbf{p})$ are the charged fermion  annihilation  and creation  operators, $\mathcal{M}$ is the amplitude of the scattering matrix, $d\mathbf{p}\equiv \frac{d^3{\bf p}}{(2\pi)^3}\frac{1}{2 p^0}$ and  $d\mathbf{q}\equiv \frac{d^3{\bf q}}{(2\pi)^3}\frac{m_f}{q^0}$ are the phase spaces of the photon and charged lepton, respectively, with similar definitions for $d\mathbf{p'}$ and $d\mathbf{q'}$.

%%%%%%%%%%%%%%%%%%%%%%%%%%%%%%%%%%%%%%        Photon- Charged lepton scattering in SME           %%%%%%%%%%%%%%%%%%%%%
\section{The generation of circular polarization in the SME}\label{sec:III}
%%%%%%%%%%%%%%%%%%%%%%%%%%%%%%%%%%%%%%%%%%%%%%%%%%%%%%%%%%%%%%%%%%%%%%%%%%%%%%%%%%%
We express  the scattering amplitude $\mathcal{M}$ of the laser beam photons and charged lepton beam in the presence of an LV background by using the minimal SME Lagrangian in the QED sector \cite{Colladay:1998fq}:
\begin{eqnarray}\label{QEDSME}
\mathcal{L}_{\text{QED}}^{\text{\tiny{LV}}}&=& \bar{\psi} [i\Gamma_\mu \mathcal{\overleftrightarrow{D}}^{\mu} -M]\psi -\frac{1}{4}F^{\mu \nu}F_{\mu \nu}-\frac{1}{4}(k_F)_{\alpha \beta\mu\nu}F^{\alpha\beta}F^{\mu\nu}+\frac{1}{2}(k_{AF})^\alpha \epsilon_{\alpha \beta\mu\nu}A^{\beta}F^{\mu\nu},
\end{eqnarray}
where $\mathcal{D}^{\mu}$ is the usual covariant derivative, $\psi$ is the charged lepton field with 
\begin{eqnarray}
M&=&m+ a_{\mu}\gamma^{\mu}-b_{\mu}\gamma^{\mu}\gamma ^5+\frac{1}{2} H_{\mu \nu}\sigma^{\mu \nu}+i m_5\gamma^5,\nonumber\\
\Gamma_{\mu}&=& \gamma_{\mu}+c_{\mu \nu}\gamma^{\nu}-d_{\mu \nu} \gamma^{\nu}\gamma^5+e_{\mu}+if_{\mu}\gamma^5+\frac{1}{2}g_{\alpha\nu\mu}\sigma^{\alpha\nu},
\end{eqnarray}
 where  $\gamma^\mu$ is the usual Dirac matrices and parameters $\{c_{\mu\nu},d_{\mu \nu},a_\mu,b_\mu,H_{\mu \nu}\}$  denote LV lepton sector coefficients and  $\{(k_F)_{\alpha \beta \mu \nu}, (k_{AF})^\alpha\}$ are the photon sector LV coefficients. Setting the coefficients equal to zero leads to the usual Dirac Lagrangian. LV coefficients emerging in M are mass-like terms and important at low energies and can easily be neglected at high energies, whereas the  parameters in $\Gamma^\mu$ which are momentum-like are considerable at high energies. In the laser and the fermion beam interaction at high energy we only are required to consider the parameters $c_{\mu\nu}$ and $d_{\mu \nu}$ in electron LV sector which are dominant. These parameters are Hermitian coefficients, dimensionless  with both symmetric and asymmetric space-time components.
\par
Considering Eqs,(\ref{eq:interactingH}) and (\ref{QEDSME}), the time evolution of the density matrix components is given as follows \cite{Bavarsad:2009hm}:
\begin{eqnarray} \label{eq:4.42}
2k^0 \frac{d} {dt} \rho_{ij}&=&\frac{e^{2}}{m_f}\int d {\bf
	q}\,n_{f}(\mathbf{x},\mathbf{q})\left[ \delta_{si}\rho_{rj}( {\bf
	k})-\delta_{rj}\rho_{is}({\bf k})
\right]
\Big[\frac{1}{2k\cdot  q}\Big\{c^{\mu\nu}\epsilon_{\mu}^{s}\Big (q\cdot k\epsilon_{\nu}^{r}-q\cdot \epsilon^{*r}\,k_{\nu}\Big) \nonumber\\&+&c^{\nu\mu}\epsilon_{\mu}^{*r}
\Big (q\cdot k\epsilon_{\nu}^{s}-q\cdot \epsilon^{s}\,k_{\nu}\Big)+
c^{\nu\mu}q_{\mu}\Big(q\cdot \epsilon^{s}
\epsilon^{*r}_{\nu}+q\cdot \epsilon^{*r}\epsilon^{s}_{\nu}\Big)\Big\}
+\frac{2c^{\nu\mu}}{(q\cdot k)^{2}} \nonumber\\&\times&
\Big (q_{\mu}q_{\nu}+k_{\mu}k_{\nu}\Big)\Big(q\cdot \epsilon^{s}\,q\cdot \epsilon^{*r}\Big)\Big],\,\,\,\,\,\,\,\,\,\,
\end{eqnarray}
where $n_{f}(\mathbf{x},\mathbf{q})$ indicates the distribution function  of charged lepton, $\ep^{s/r}_{\mu}(k)$s are the photon polarization states with $s,r=1,2$.
The energy density $\varepsilon_f(\mathbf{x})$, the number density $n_f(x)$ and the averaged momentum $\bar{q}$ of the leptons are defined as
\begin{eqnarray}\label{nd-n0}
\frac{ n_f(\mathbf{x})}{\bar{q}}&=&\int \frac{d^3 \bf{q}}{(2 \pi)^3}~\frac{n_f(\mathbf{x},\mathbf{q})}{q},\,\,\,n_f(\mathbf{x})=\int \frac{d^3 \bf{q}}{(2 \pi)^3} n_f(\mathbf{x},\mathbf{q}),\nonumber\\ \varepsilon_f(\mathbf{x})&=&g_{f}\int \frac{d^3 \bf{q}}{(2\pi)^3}~q^0 \,n_f(\mathbf{x},\mathbf{q}),
\end{eqnarray}
where $g_f$ is the number of spin states. In the calculation, we assume $n_f(\mathbf{x},\bar{\mathbf{q}})\sim \exp[-|\mathbf{q}-\bar{\mathbf{q}}|/|\bar{\mathbf{q}}|]$ for charged lepton beam, which means that  most of the leptons are moving with the  same momentum $\mathbf{q}\simeq \bar{\mathbf{q}}$ and in the same direction.
We should note that at forward photon-lepton scattering, the term including the $d_{\mu \nu}$ correction does not generate any circular polarization, i.e. $ \dot V=0$ 
\cite{Bavarsad:2009hm}. However, $d_{\mu \nu}$ may contribute to circular polarization in a higher order correction, which is proportional to $(d_{\mu\nu})^2$ and negligible. Hence, $c_{\mu \nu}$ is the only source of LV which we will consider below. \par
From many phenomenological points of view, it is more convenient to consider $c_{\mu \nu}$ as a symmetric and traceless tensor, therefore having nine independent components for the $c$ coefficients. In fact, the anti-symmetric part at leading order is equivalent to the redefinition in the representation of the Dirac matrices \cite{Altschul:2006pv}. Therefore the physical quantities are independent of the anti-symmetric part of the c coefficient at leading order.
Then the time evolution of the Stokes parameters given in Eq. (\ref{eq:density matrix}) is obtained as follows: \cite{Bavarsad:2009hm}
\begin{eqnarray}\label{eq:Stoks-time}
\dot I(k)&=&0\,\,,\,\,\,\,\,\,\,\dot V(k)=\rho_{Q}\,Q(k)+\rho_{U}\,U(k) \nonumber\\
\dot Q(k)&=&-\rho_{Q}\,V(k)\,\,,\,\,\,\,\,\,\,\dot U(k)=-\rho_{U}\,V(k).
\end{eqnarray}
Solving the above coupled equations leads to
\begin{equation}\label{eq:differentialequation}
\ddot V=\dot \rho _{Q} Q+\dot \rho _{U} U-\rho^2 V,
\end{equation} 
with
\begin{eqnarray}\label{rhos}
\rho_Q&=&-\frac{ie^{2}}{2k^{0}m}\int {d{\bf
		q}}~n_f(\mathbf{x},\mathbf{q}) \, \frac{c^{S\mu\nu}}{4(q\cdot  k)^2} \Big[2( q\cdot k)^2(\epsilon_{\mu}^{2}\epsilon_{\nu}^{1}+\epsilon_{\mu}^{1}\epsilon_{\nu}^{2})+q \cdot k\,\Big(q\cdot  \epsilon^{2}\big((q_{\mu}-k_{\mu})\epsilon_{\nu}^{1}\nonumber\\&+&(q_{\nu}-k_{\nu})\epsilon_{\mu}^{1}\big)+q\cdot  \epsilon^{1}\big((q_{\mu}-k_{\mu})\epsilon_{\nu}^{2}+(q_{\nu}-k_{\nu})\epsilon_{\mu}^{2}\big)\Big)+
8\,(q\cdot  \epsilon^{1}q.\epsilon^{2})(q_{\nu}q_{\mu}+k_{\nu}k_{\mu})
\Big],\nonumber\\
\rho_U&=&\frac{ie^{2}}{2k^{0}m}\int {d{\bf
		q}}~n_f(\mathbf{x},\mathbf{q}) \, \frac{c^{S\mu\nu}}{4(q\cdot  k)^2} \Big[2( q\cdot k)^2(\epsilon_{\mu}^{1}\epsilon_{\nu}^{1}-\epsilon_{\mu}^{2}\epsilon_{\nu}^{2})+q \cdot k\,\Big(q\cdot  \epsilon^{2}\big((k_{\mu}-q_{\mu})\epsilon_{\nu}^{2}\nonumber\\&+&(k_{\nu}-q_{\nu})\epsilon_{\mu}^{2}\big)+q\cdot  \epsilon^{1}\big((q_{\mu}-k_{\mu})\epsilon_{\nu}^{1}+(q_{\nu}-k_{\nu})\epsilon_{\mu}^{1}\big)\Big)+
4\,(q\cdot  \epsilon^{1}q\cdot\epsilon^{1}-q\cdot \epsilon^{2}q\cdot \epsilon^{2})(q_{\nu}q_{\mu}+k_{\nu}k_{\mu})
\Big]
\end{eqnarray}
and $\rho=\sqrt{\rho^{2} _{Q} +\rho^{2} _{U}}$, $c^{S \mu \nu}$ means the symmetric part of $c^{\mu \nu}$ tensor, $q$ and $k$ are the momenta of leptons and photons, respectively.
In particular, available bounds on coefficients of  the SME  are given in the Standard Sun-centered non-rotating inertial reference frame. However, the experimental set up is usually managed based on the Earth. Therefore, for parameterizing our result, it is necessary to introduce Cartesian coordinates on the Earth frame and suitable basis of vectors for a non-rotating frame. Let us define a coordinate system in three-dimensional space. $(\hat X,\hat Y,\hat Z)$ shows the non-rotating basis in which $\hat Z$ points along the Earth's axis on the north direction. 
Then the non-relativistic transformation to a lab basis $(\hat x,\hat y,\hat z)$  at any time $t$ is given by \cite{Kostelecky:1999mr}
\begin{equation}
\begin{pmatrix}  \hat{x}\cr \hat y\cr \hat z\cr\end{pmatrix}
=\begin{pmatrix} \cos \chi \cos {\Omega t}&
\cos \chi \sin {\Omega t}& -\sin \chi\cr  -\sin {\Omega t} & \cos{\Omega t} &0\cr
\sin \chi \cos {\Omega t}&
\sin \chi \sin {\Omega t}& \cos \chi\cr \end{pmatrix}
\begin{pmatrix} \hat X\cr \hat Y\cr \hat Z\cr \end{pmatrix} ,
\end{equation}
where  $\Omega\simeq 2\pi/(23$h 56 min) is the sidereal rotation frequency of the Earth and $\chi=\hat Z\cdot\hat z$ varies as $0\le\chi\le \pi$.  In order to specify the direction of Earth-based frame we identify the equator and north pole as $\chi=\pi/2$ and $\chi=0$, respectively. When $\chi=0,$ then $\hat {Z}||\hat{z}$ and hence $\hat{z}$ indicates a normal to the surface of  Earth, when $\chi=\pi/2$, the $\hat{x}$ is anti-parallel to $\hat {Z}$. Therefore $\hat{x}$ and $\hat{y}$ point south and east, respectively. The time is chosen in such a way that $\hat{z}(t=0)$ stays in the quadrant of the $\hat{Z}-\hat{X}$ surface. Since the lab frame rotates with rotation of the Earth, the LV components on the Earth will depend on location and time i.e. the experimental observable would vary with time and location.  By transforming the time-like and space-like vectors of the tensor $c_{\mu
    \nu}$ we obtain
\begin{eqnarray}
c_{00}&=& c_{TT},\nonumber\\ 
c_{22}&=&\cos ^2\chi  \big(c_{XX} \cos ^2 \Omega t+(c_{XY}+c_{YX}) \sin \Omega t  \cos \Omega t+c_{YY }\sin ^2 \Omega t \big)-\sin \chi  \cos \chi (c_{XZ}+c_{ZX}) \cos \Omega t\nonumber\\&+&(c_{YZ}+c_{ZY}) \sin
\Omega t)+c_{ZZ} \sin ^2\chi ,\nonumber\\ 
c_{33}&=&c_{XX} \sin ^2(\Omega t)-(c_{XY}+c_{YX}) \sin\Omega t \cos \Omega t+c_{YY} \cos ^2 \Omega t,\nonumber\\ 
c_{(01)}&=&\cos \chi ((c_{TX}+c_{XT}) \cos \Omega t+(c_{TY}+c_{YT}) \sin \Omega t)-(c_{TZ}+c_{ZT}) \sin \chi,\nonumber\\ 
c_{(02)}&=&(c_{TY}+c_{YT}) \cos \Omega t-(c_{TX}+c_{XT}) \sin\Omega t,\nonumber\\
c_{(03)}&=&\sin \chi (c_{TX}+c_{XT}) \cos \Omega t+(c_{TY}+c_{YT}) \sin \Omega t)+(c_{TZ}+c_{ZT}) \cos \chi,\nonumber\\ 
c_{(12)}&=&\cos \chi  ((c_{YY}-c_{XX}) \sin  2 \Omega t+(c_{XY}+c_{YX}) \cos 2 \Omega t)+\sin \chi  ((c_{XZ}+c_{ZX}) \sin \Omega t-(c_{YZ}+c_{ZY}) \cos \Omega t),\nonumber\\ 
c_{(13)}&=& \sin 2\chi\left(c_{XX} \cos^2\Omega t+(c_{XY}+c_{YX}) \sin \Omega t \cos \Omega
	t+c_{YY} \sin ^2\Omega t-c_{ZZ}\right)+\cos ^2\chi (c_{XZ} \cos \Omega t+c_{YZ} \sin \Omega
	t)\nonumber\\
	&-&\sin ^2\chi (c_{XZ} \cos \Omega t+c_{YZ} \sin\Omega t)+\cos ^2\chi (c_{ZX} \cos \Omega t+c_{ZY} \sin \Omega t)-\sin ^2 \chi(c_{ZX} \cos \Omega t +c_{ZY} \sin \Omega t),\nonumber\\ 
c_{(23)}&=&\sin \chi  ((c_{YY}-c_{XX}) \sin 2 \Omega t+(c_{XY}+c_{YX}) \cos 2 \Omega t)+\cos \chi 
((c_{YZ}+c_{ZY}) \cos\Omega t-(c_{XZ}+c_{ZX}) \sin\Omega t),
\label{c-coefficients}
\end{eqnarray}
where $c_{(ij)}=c_{ij}+c_{ji}$ with $i,j=0,...,3$. The time dependence of the LV parameter causes a day-night asymmetry in FC. However, the interaction of a photon with the lepton beam occurs at a very short time of order $\Delta_{t_I} \approx 10^{-15}-10^{-14}$s which we will show in the following. At this period of time, rotation of Earth is negligible and the components of $c_{\mu \nu}$  are not changing during this time. However, as seen from Eq. (\ref{c-coefficients}), the time of performing the experiment during the day-night might have impact on the experimental results.% so the asymmetry is not needed to be considered. 
	\begin{table}
		\centering
		\caption{\small{ Experimental bounds on components of c.}}
		\vspace{0.3cm}
		\scalebox{0.95}
		{ \begin{tabular}{lll}
				\hline\hline
				$c_{\mu \nu} $&~Experimental bounds&~~~~~~~~~ system\\
				\hline
				$c_{TT}$&$~~~~~~~~~2\times 10^{-15}$&~~ Collider physics \cite{Altschul:2010na}\\
				% \hline
				%$c_{XX}$&$~~~~~~~~~5\times 10^{-15}$&~~~Astrophysics \cite{Altschul:2006pv}\\
				\hline
				$c_{YY}$&$~~~~~~~~~3\times 10^{-15}$&~~~Astrophysics \cite{Altschul:2006pv}\\
				\hline
				$c_{ZZ}$&$~~~~~~~~~5\times 10^{-15}$&~~~Astrophysics \cite{Altschul:2006pv}\\
				\hline
				$c_{(XY)}$&$~~~~~~~~~3\times 10^{-15}$&~~~Astrophysics \cite{Altschul:2006pv}\\
				\hline
				$c_{(YZ)}$&$~~~~~~~~~1.8\times 10^{-15}$&~~~Astrophysics \cite{Altschul:2006pv}\\
				\hline
				$c_{(XZ)}$&$~~~~~~~~~3\times 10^{-15}$&~~~Astrophysics \cite{Altschul:2006pv}\\
				\hline
				$c_{(TX)}$&$~~~~~~-30\times 10^{-14}$&~~~Collider physics \cite{Altschul:2010na}\\
				\hline
				$c_{(TY)}$&$~~~~~~-80\times 10^{-15}$&~~~Collider physics \cite{Altschul:2010na}\\
				\hline
				$c_{(TZ)}$&$~~~~~~-11\times 10^{-13}$&~~~Collider physics \cite{Altschul:2010na}\\
				\hline\hline \\
			\end{tabular}} \label{tab:Conversion1}
		\end{table}

	As explained above the $\rho_{Q}$ and $\rho_{U}$ given in Eq(\ref{rhos}) are almost independent of interaction time. Therefore, the above differential equations (Eq(\ref{eq:differentialequation})) reduce to
	\begin{equation}\label{eq:Dif}
	\ddot V+\rho ^{2}V=0.
	\end{equation}
	The general solution for above equation is
	\begin{align}\label{a39}
	V(t_I)=\mathcal{A}\sin(\rho~ t_I)+\mathcal{B}\cos(\rho~ t_I);
	\end{align}
	On applying the initial condition at $t_{I}=0$, $V(0)=0$, for a totally linear polarized laser beam, Eq. (\ref{eq:FC}) can be rewritten as follows:
	\begin{eqnarray}\label{eq:vdot}
	\Delta \phi_{FC}=\frac{\rho_Q\, {Q_0}+\rho_U \,{U_0}}{2\rho}\, \sin\big(\rho\,\Delta t_{I}\big),
	\end{eqnarray}
	where $\Delta t_{I}$ is the time that the beams are interacting. Considering Eq(\ref{rhos}), we obtain
	\begin{equation}\label{phi-final}
	\Delta \phi_{FC}\sim \mathcal{N} \sin\big(\frac{3}{16}\frac{m_e^2\, \sigma_T}{\alpha\, k_0\, q_0}\sqrt{(w_Q^2+w_U^2)}\frac{\bar \ep_i(\mathbf{x},\bar{\mathbf{q}})}{q_0}\,\Delta t_{I}\big),
	\end{equation}
	with
	\begin{eqnarray}\label{wqu}
	w_U &=& -2\,\frac{ (q\cdot\epsilon^{1}q\cdot\epsilon^{1}-q\cdot\epsilon^{2}q\cdot\epsilon^{2})(q\cdot c^S\cdot q +k\cdot c^S\cdot k)}{(q\cdot k)^2},\nonumber\\
	w_Q &=& 4\,\frac{ q\cdot\epsilon^{1}q\cdot\epsilon^{2}(q\cdot c^S\cdot q +k\cdot c^S\cdot k)}{(q\cdot k)^2}.
	\end{eqnarray}
	Here we should note that in deriving the above equations, we only consider dominant terms in  Eq. (\ref{rhos}),  and the normalization factor is $\mathcal{N}=\frac{w_Q Q_0+w_U U_0}{\sqrt{w_U^2+w_Q^2}}$. 

	%%%%%%%%%%%%%%%%%%%%%%%%%%%%%%%%%%%%%%%%%%%%%%%%%%%%%%%%%%%%%%%%%%%%%%%%%%%%%%%%%%%%%%%%%%%%
	\section{ Set up of laser and charged beams collision}\label{sec:IV}
	%%%%%%%%%%%%%%%%%%%%%%%%%%%%%%%%%%%%%%%%%%%%%%%%%%%%%%%%%%%%%%%%%%%%%%%%%%%%%%%%%%%%%%%%%%%%
	In the lab frame,  we  set the direction $\hat{{ k}}$ and polarization vectors $\vec\epsilon_s(k)$ of the incident laser beam photons in the observer's Minkowskian frame coordinate system as follows:
	\begin{eqnarray}
	\hat{{ k}}&=&(\sin\theta \cos\varphi ,\sin\theta \sin\varphi, \cos\theta),\nonumber\\
	\vec\epsilon_1(k)&=&(\cos\theta \cos\varphi, \cos\theta \sin\varphi, -\sin\theta),\nonumber\\
	\vec \epsilon_2(k)&=&(-\sin\varphi, \cos\varphi,0),
	\end{eqnarray}
	in which $\hat k=\frac{\vec{k}}{|k|}$ makes an angle of $\theta$ with the $z$-axis
	and the 4-vector momentum of charged lepton $q$
	has been assumed to be in the $\hat{z}$-direction.\par
	There are some charged lepton beams with different energies and luminosity. The most appropriate charged lepton beams for our purpose are the future muon and electron beams available at colliders such as multi-TeV muon collider, Muon Accelerator Program (MAP) \cite{{Bravar:2011dx},{Delahaye:2013jla}},  electron-positron International Linear Collider (ILC) \cite{Barklow:2015tja} and Compact Linear Collider (CLIC), which is planned to obtain a center of mass energy up to $\sqrt{s}= 3$ TeV \cite{{CLIC:2016zwp},{Fujii:2015jha}}.\par
	 \begin{figure}
	 	\includegraphics[width=0.43\linewidth]{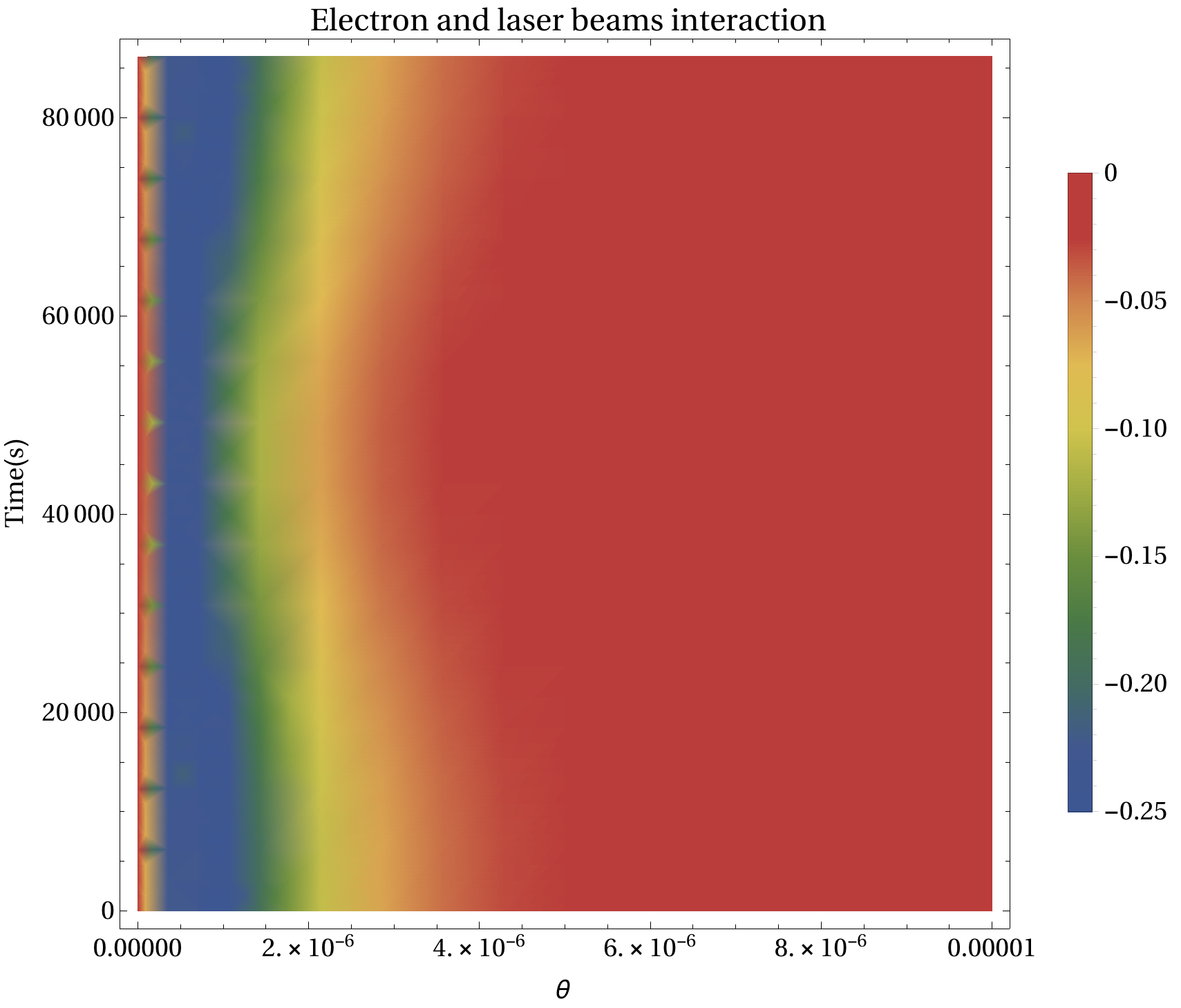}
	 	\includegraphics[width=0.43\linewidth]{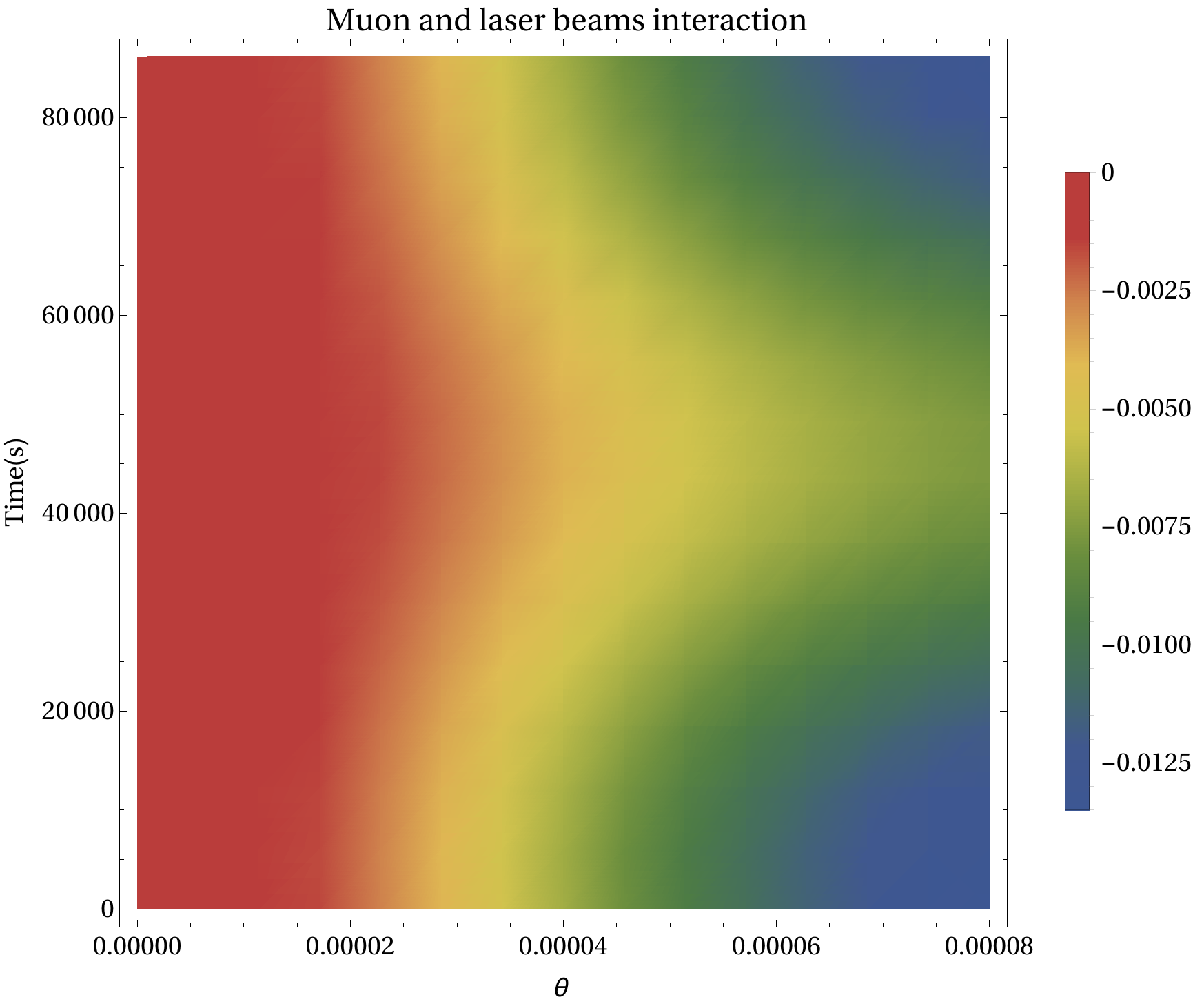}
	 	\hspace{1cm}
	 	\caption{\small FC phase $\Delta \phi|^{\text{LV}}$ ({\bf left panel}) for the electron beam ({\bf right panel}) muon beam, interacting with the laser beam  in the time versus $\theta$ plane with $\varphi=\chi=\frac{\pi}{4}$.}
	 	\label{fig:deltaphitot}
	 \end{figure}
    The basic picture of our experiment is fairly simple. We measure the generation of circular polarization for the laser beam via its forward scattering with the charged lepton beam in the presence of LV effects. Equation (\ref{phi-final}) shows that in order to generate a wide range of circularly polarized photons, the intensity of the lepton  and laser beam should be large enough. However, in high-intensity laser, the effects of the magnetic part of the Lorentz force on the charged lepton become important. In this case, the magnetic field as a trivial background will produce circular polarization\cite{{Cooray:2002nm},{Bavarsad:2009hm}}. Moreover, at strong electromagnetic fields, the nonlinear  Compton scattering effect become significant \cite{Harvey:2009ry}. In addition to the collision of the laser beam with an electron high-energy beam, photons would possibly obtain high energy by back-scattering off the high-energy electron beam, i.e. we have inverse Compton scattering. Therefore, pair production is possible as a result of Breit-Wheeler reaction. However, the mentioned effects using a low-intensity laser beam are negligible and can safely be ignored. Another point is that low-intensity lasers are available sources of monochromatic radiation. They are lightweight and low-cost devices.
     \par
    While electromagnetic fields and the coherent time duration of the laser pulse increase, the number of laser photons colliding with an electron beam will increase as well. Moreover, forward scattering of linearly polarized photons on polarized lepton such as the electron can cause the circular polarization of photons \cite{Mohammadi:2013ksa}. To reduce unwanted backgrounds effects as explained above, which are particularly important at high frequencies and intensities, we assume the lepton beam is nearly unpolarized and the laser beam has a low intensity and energy. As shown in Eq. (\ref{eq:Stoks-time}), in addition to FC, FR also results by forward scattering in the presence of LV effects.  \par
        We consider a typical relativistic electron beam  with energy of order of $E_e\sim\mathcal{O}$ (TeV), the number of  electron per bunch is $n_e\sim\mathcal{O}(10^{10} \text{cm}^{-3})$ and the size of beam bunch is of the order $\sim\mathcal{O}(\upmu m)$.  The average energy of flux per bunch can be estimated as \cite{Tizchang:2016hml}
	\begin{equation}\label{Edensitye}
	\bar \ep_e(\mathbf{x},\bar{\mathbf{q}})\approx |\bar{\mathbf{q}}|\,n_e(\mathbf{x},\bar{\mathbf{q}})c\sim 10^{10} \,{\rm TeV}/({\rm cm}^2{\rm s}),
	\end{equation}
	which is normalized to the number density of the electrons $n_e$. The interacting time $\Delta t_{I}$ can be obtained by taking into account the size of two beams at the interacting point as $\Delta d \simeq c~ \Delta t_{I}$, where $\Delta d$ is the spatial interval of the interacting spot. 
    It means that the beam with a larger size would be more effective for our aim. Therefore, the interaction time duration for laser and electron beam is about $\Delta t_{I}\sim 10^{-15}~ $s. \par
  Furthermore, we  choose the conventional relativistic muon beam  with energy order of $E_\mu\sim\mathcal{O}$(TeV) and the number of muon per bunch is supposed to be about $n_\mu\sim\mathcal{O}(10^{12}\text{cm}^{-3})$. The size of beam bunch is $\sim\mathcal{O}(10\, \mu m)$ with the interacting time  $\Delta t_I\simeq \Delta d/c \sim 10^{-14}~ $s. the average energy of flux per bunch is given  as follows:
	\begin{equation}\label{Edensitye}
	\bar \ep_\mu(\mathbf{x},\bar{\mathbf{q}})\approx |\bar{\mathbf{q}}|\,n_\mu(\mathbf{x},\bar{\mathbf{q}})c\sim 10^{12} \,{\rm TeV}/({\rm cm}^2{\rm s}).
	\end{equation}
    		\begin{figure}[t]
    	\hspace{-1cm}
    		\includegraphics[width=0.3297 \linewidth]{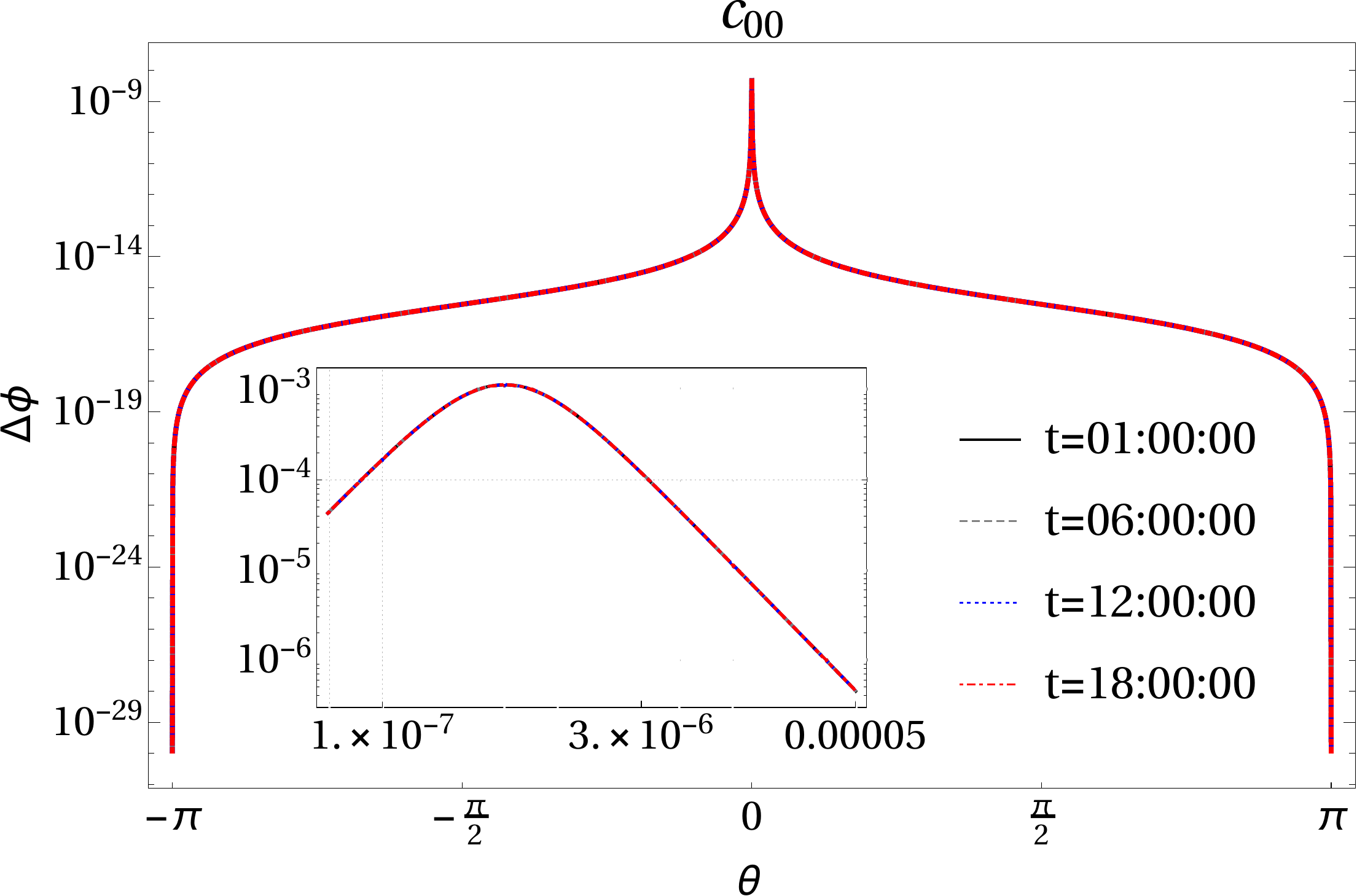}
       		\includegraphics[width=0.3297 \linewidth]{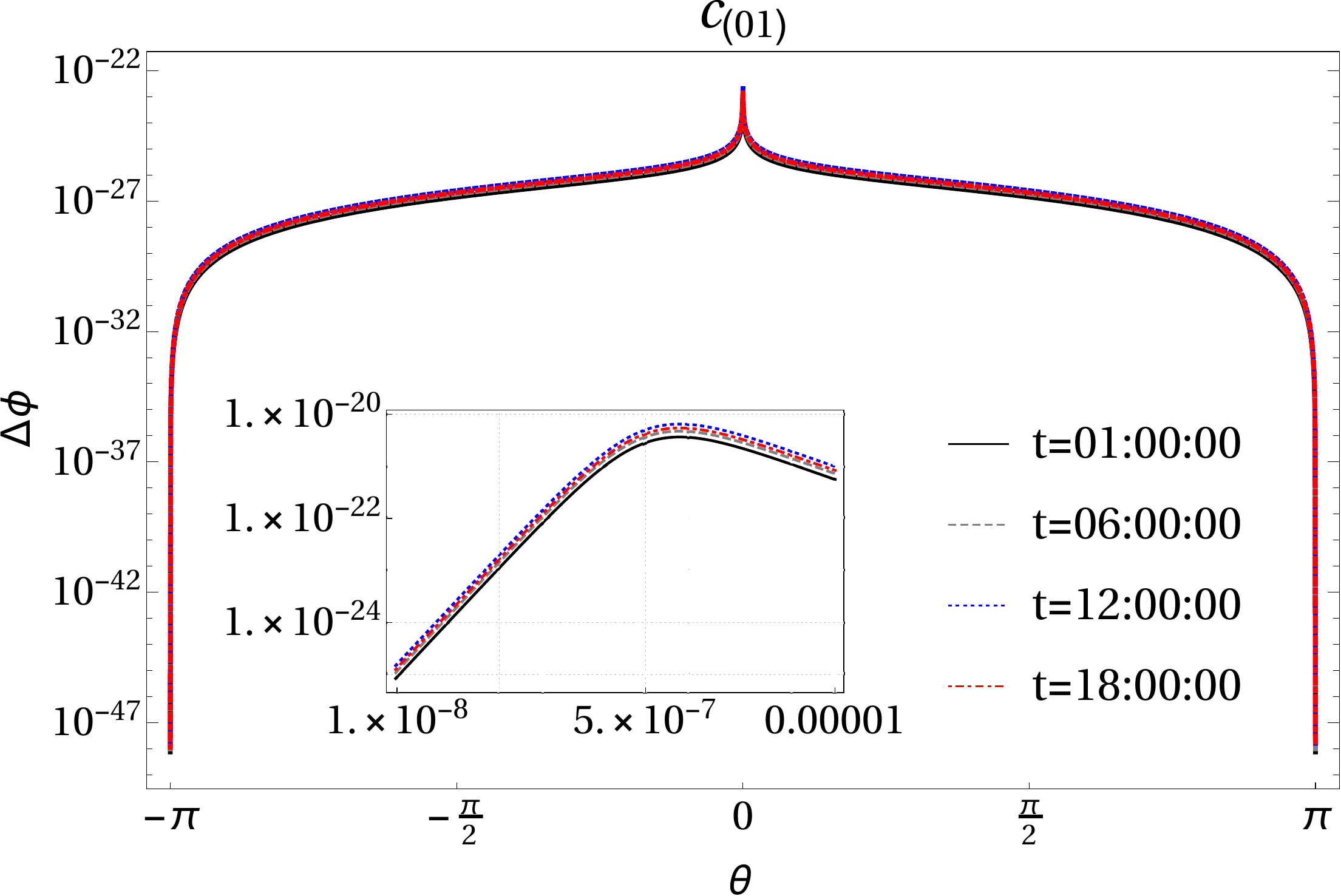}
    		\includegraphics[width=0.3297 \linewidth]{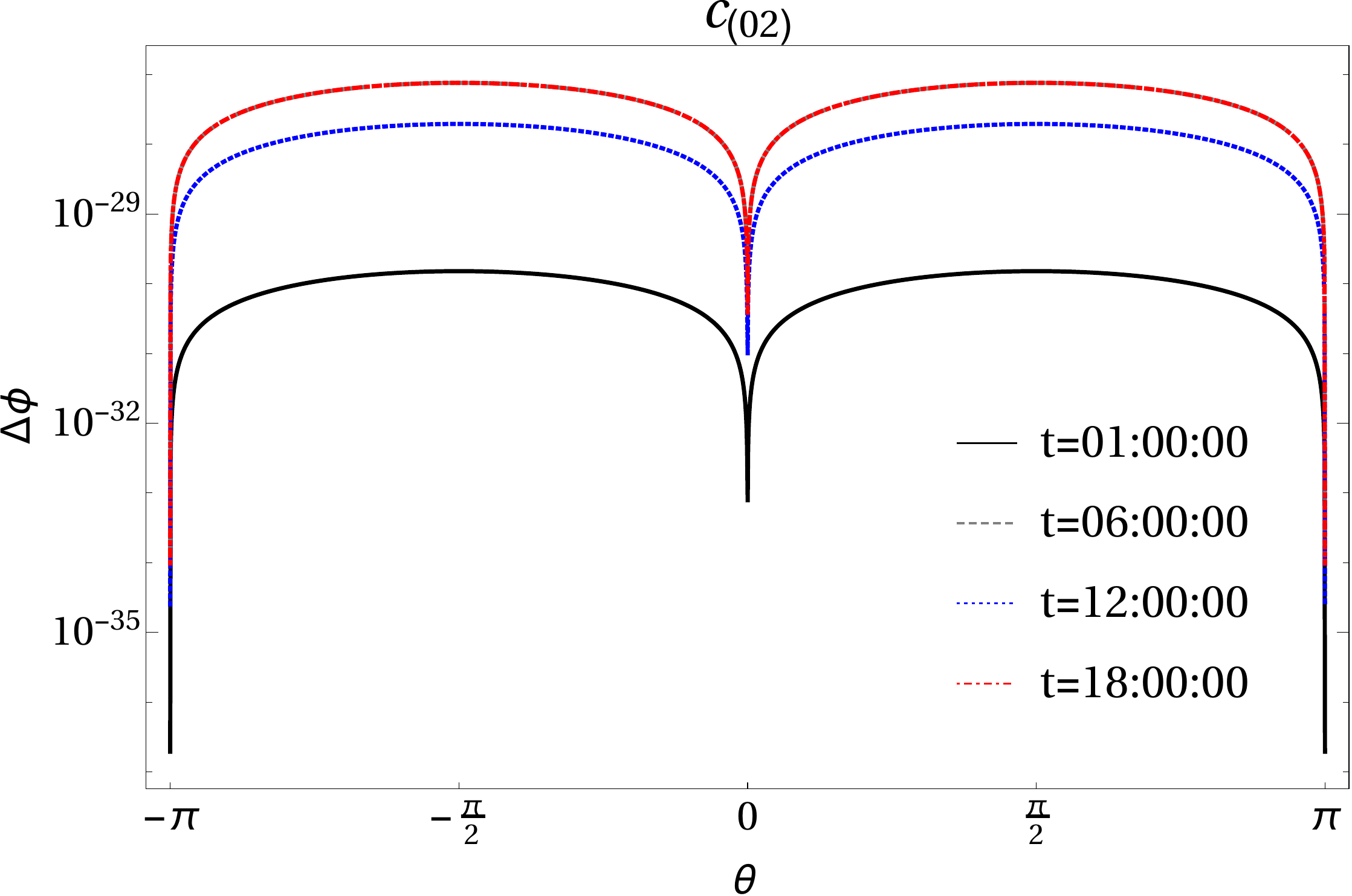}\\
    		\hspace{-1cm}
    		\includegraphics[width=0.3297 \linewidth]{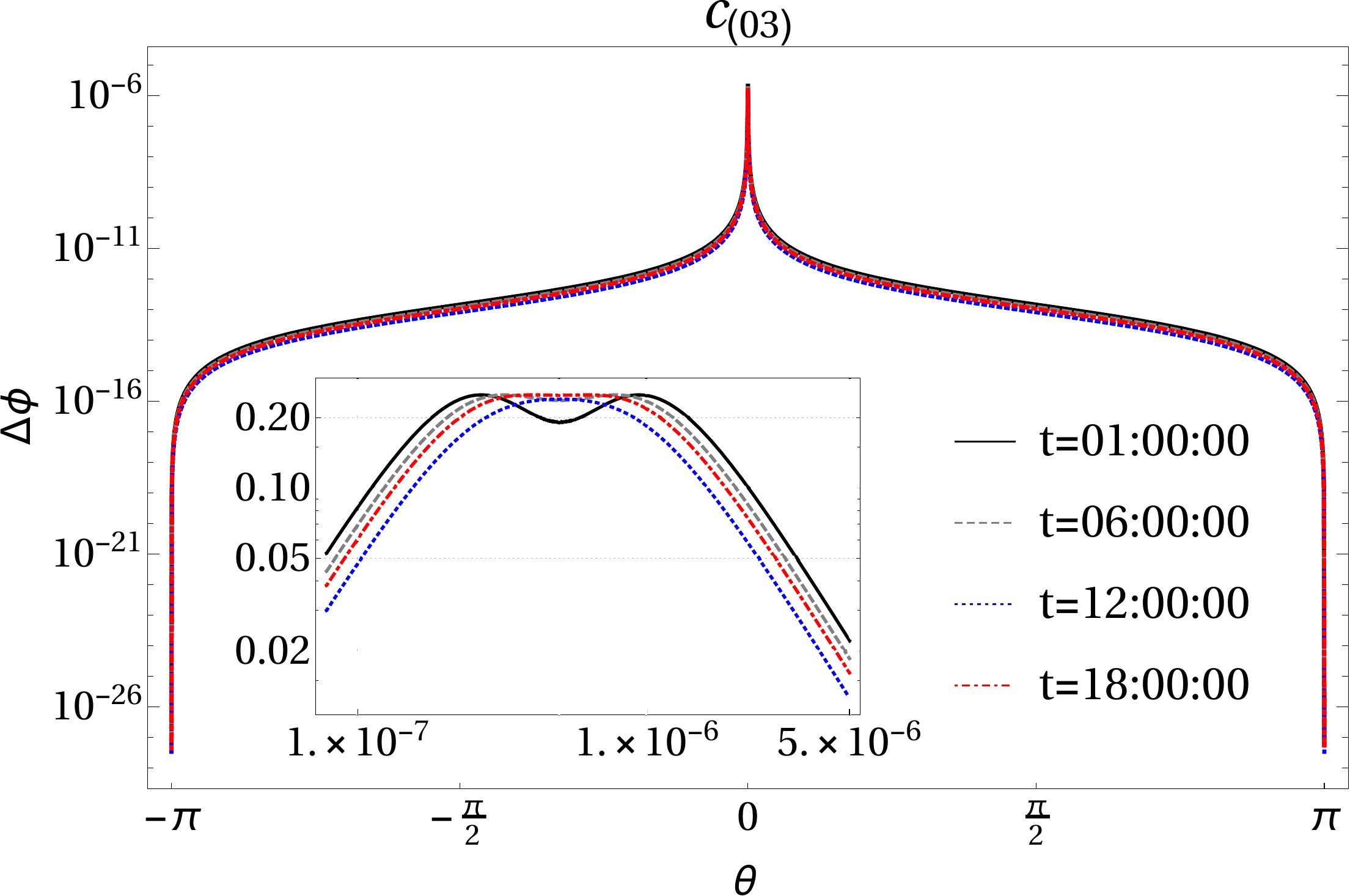}
       		\includegraphics[width=0.3297 \linewidth]{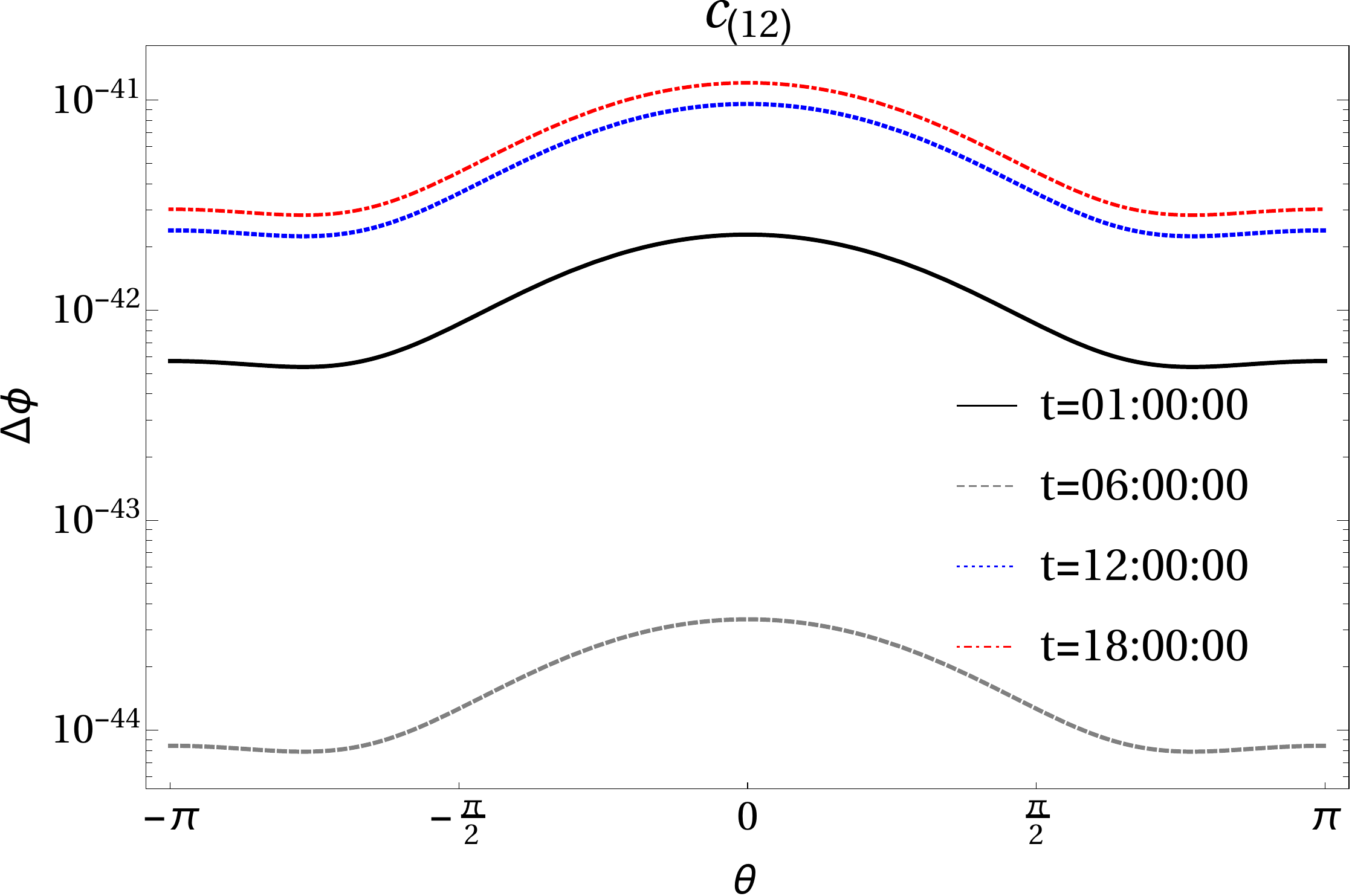}
    		\includegraphics[width=0.3297 \linewidth]{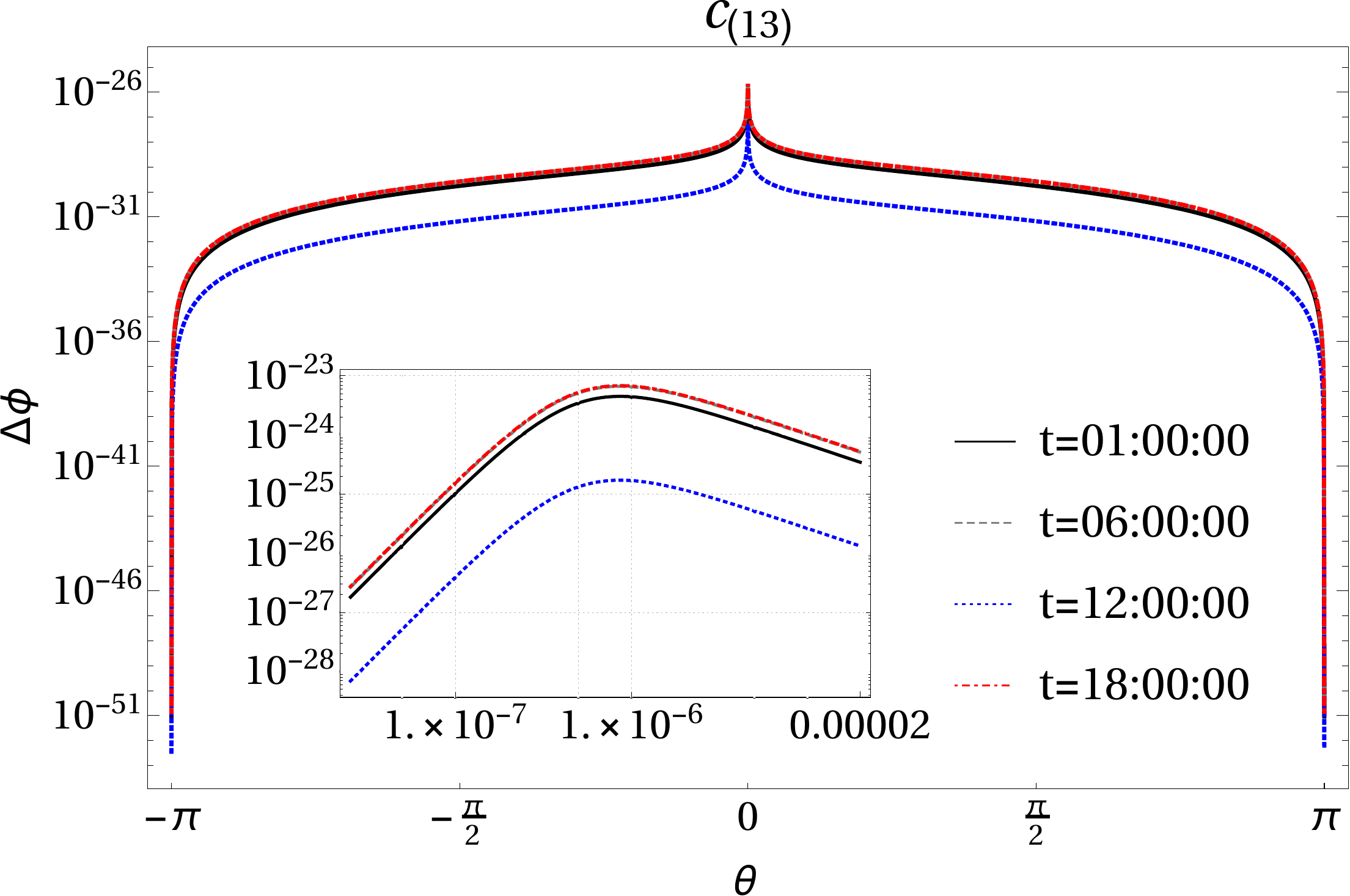}\\
    			\hspace{-1cm}
    		\includegraphics[width=0.3297 \linewidth]{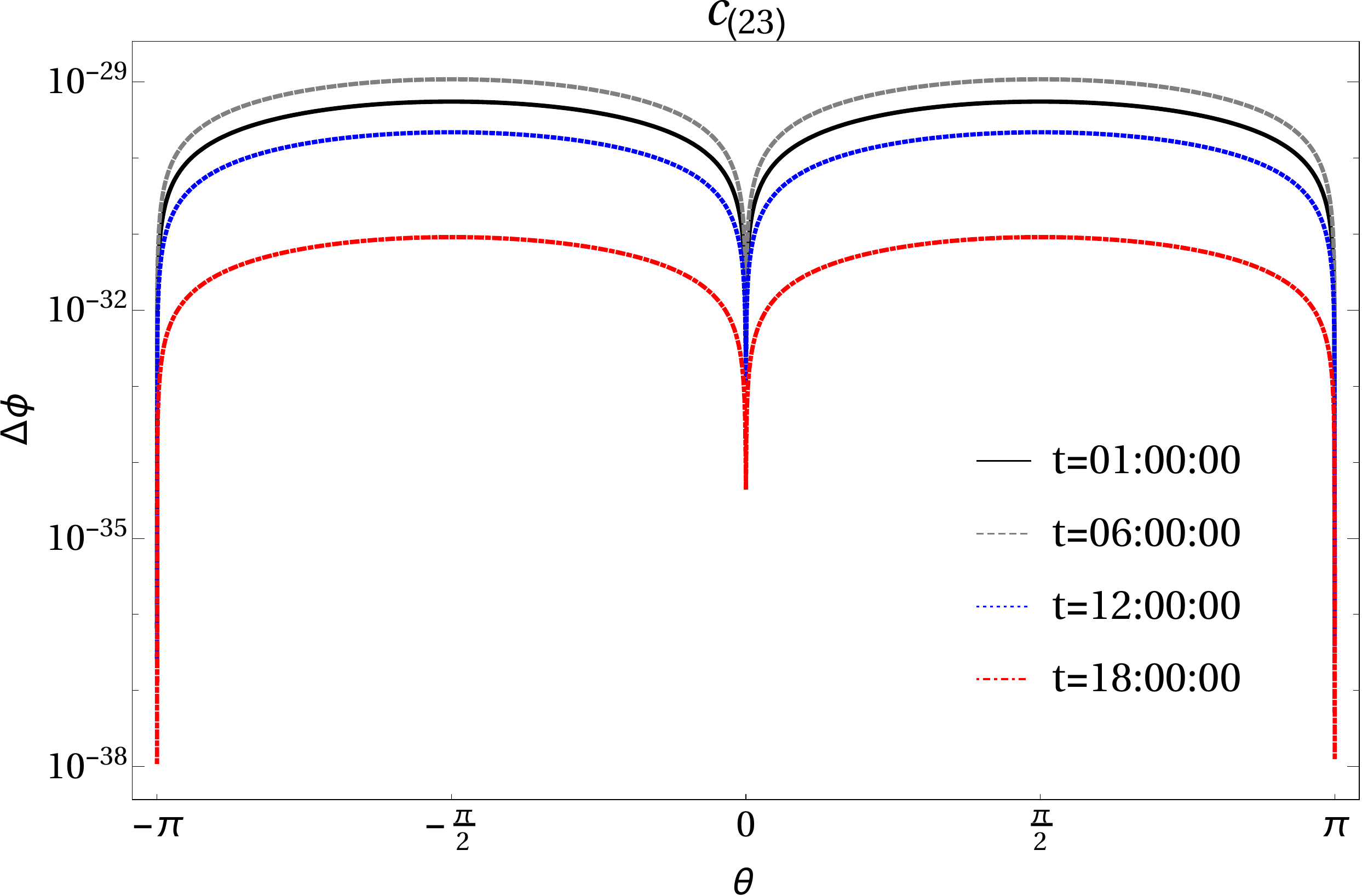}
    		\includegraphics[width=0.3297 \linewidth]{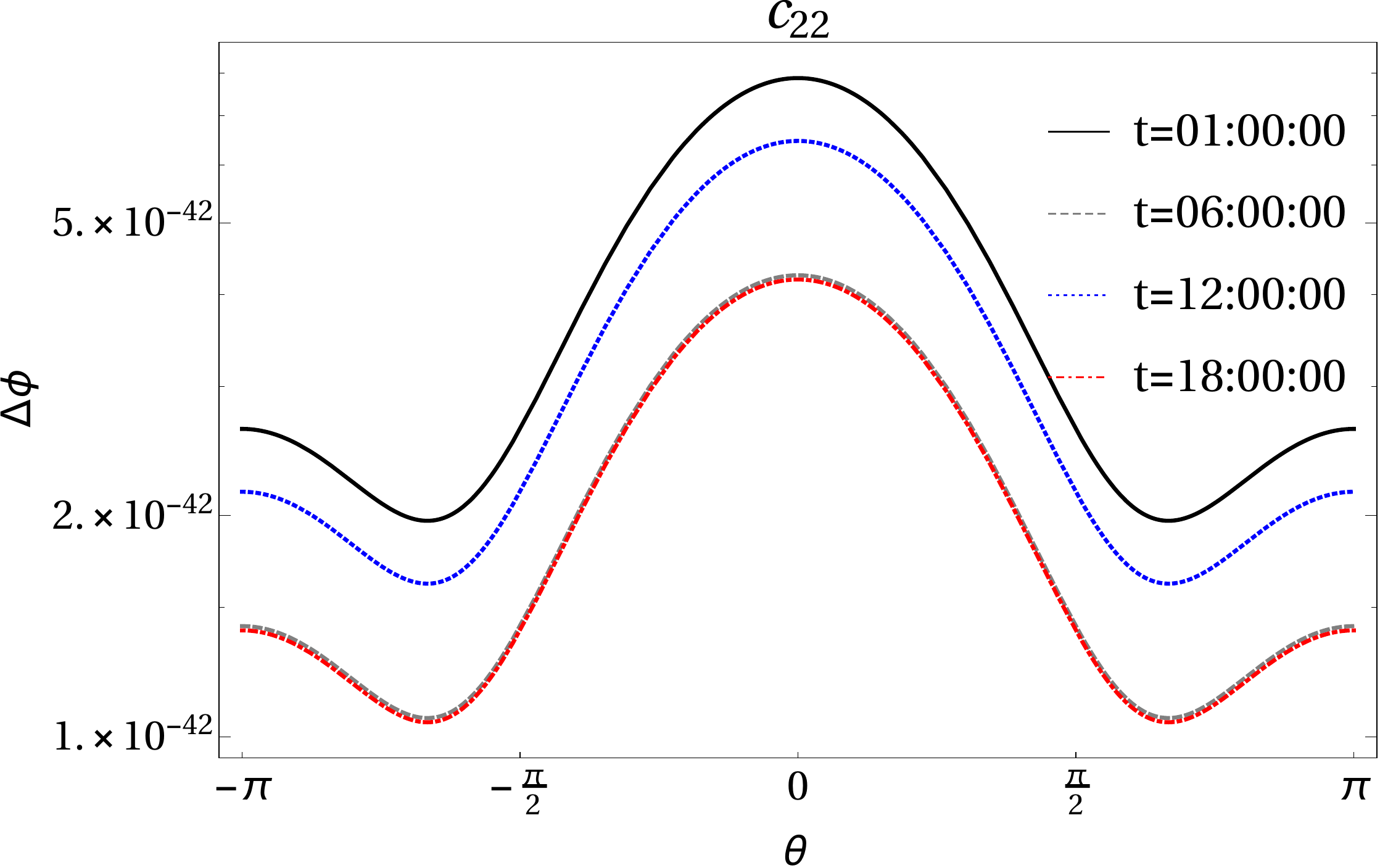}
    		\includegraphics[width=0.3297 \linewidth]{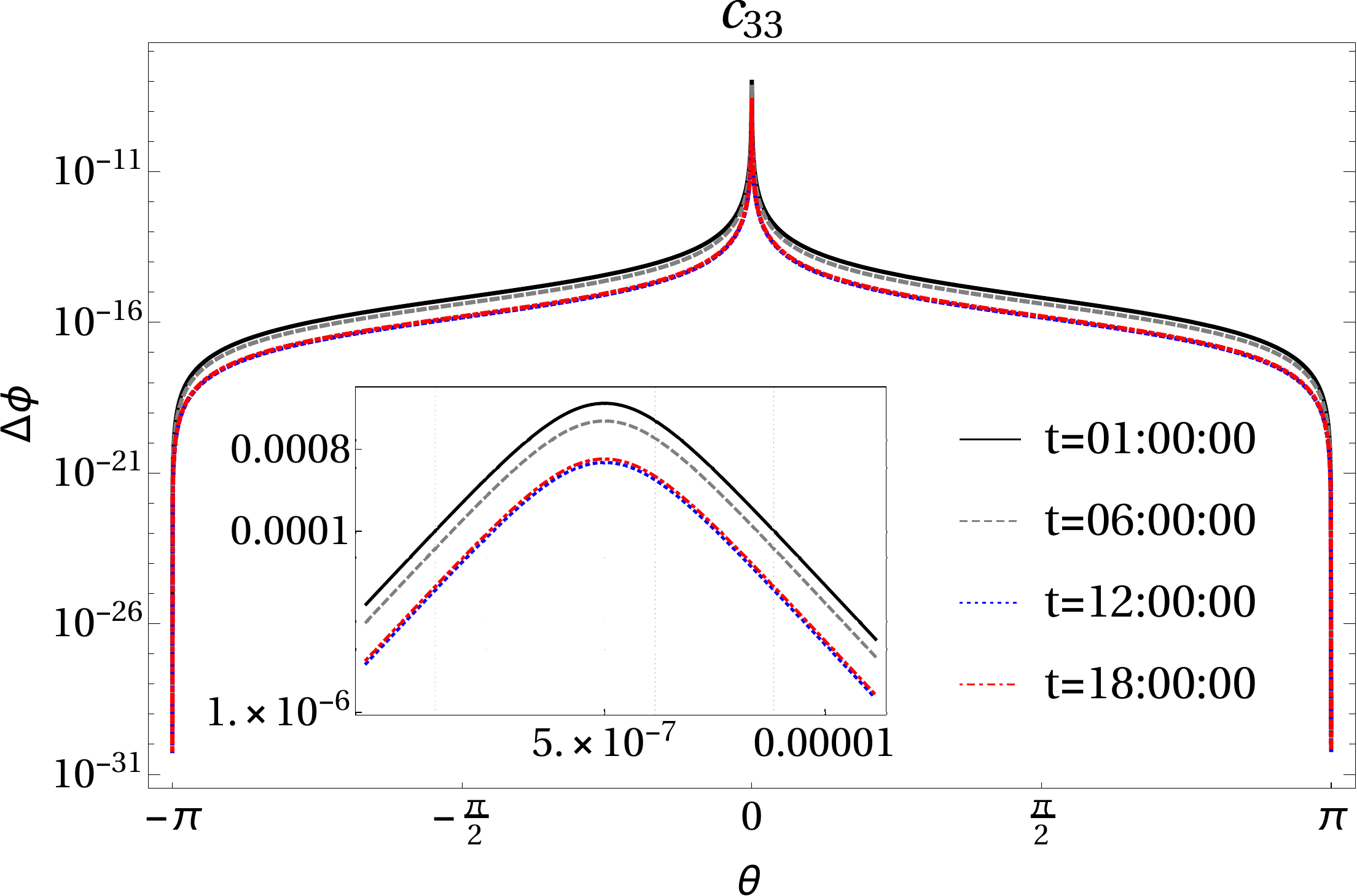}
    		\caption{\small FC phase, $\Delta \phi|_e^{\text{LV}}$ as a function of scattering angle $\theta$ for four different times during a day-night. We have taken $\varphi=\chi=\frac{\pi}{4}$.}
    		\label{fig:deltaphie}
    	\end{figure}  
    The LV parameters are broadly examined and there are constraints on each individual component or combinations of some parameters. Data-table for available upper bounds on components of $c_{\mu\nu}$ in Sun-centered reference frame achieved through some experiments is given in Tab.\ref{tab:Conversion1}. Considering these bounds and applying them to Eq. (\ref{phi-final}), we display in Fig.\ref{fig:deltaphitot} the accessible order of FC $\Delta\phi|^{\text{LV}}$  in the time of day-night versus scattering angle $\theta$ plane for both electron and muon beams interacting with the laser beam. It can be seen that the magnitude of FC for different times of the day-night is varying a little.
    		\begin{figure}[t]
    			\hspace{-1cm}
    			\includegraphics[width=0.3297 \linewidth]{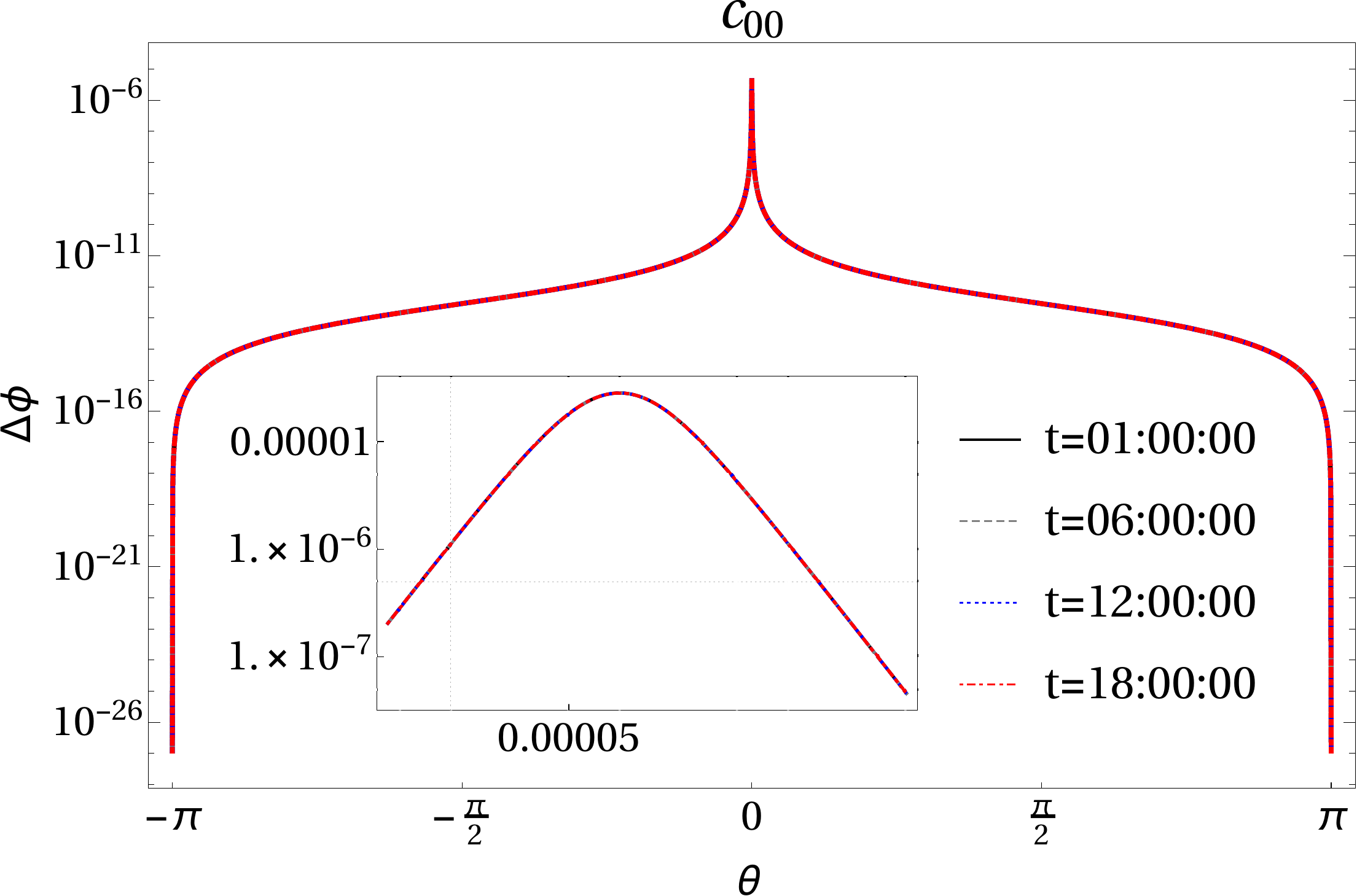}
    			\includegraphics[width=0.3297 \linewidth]{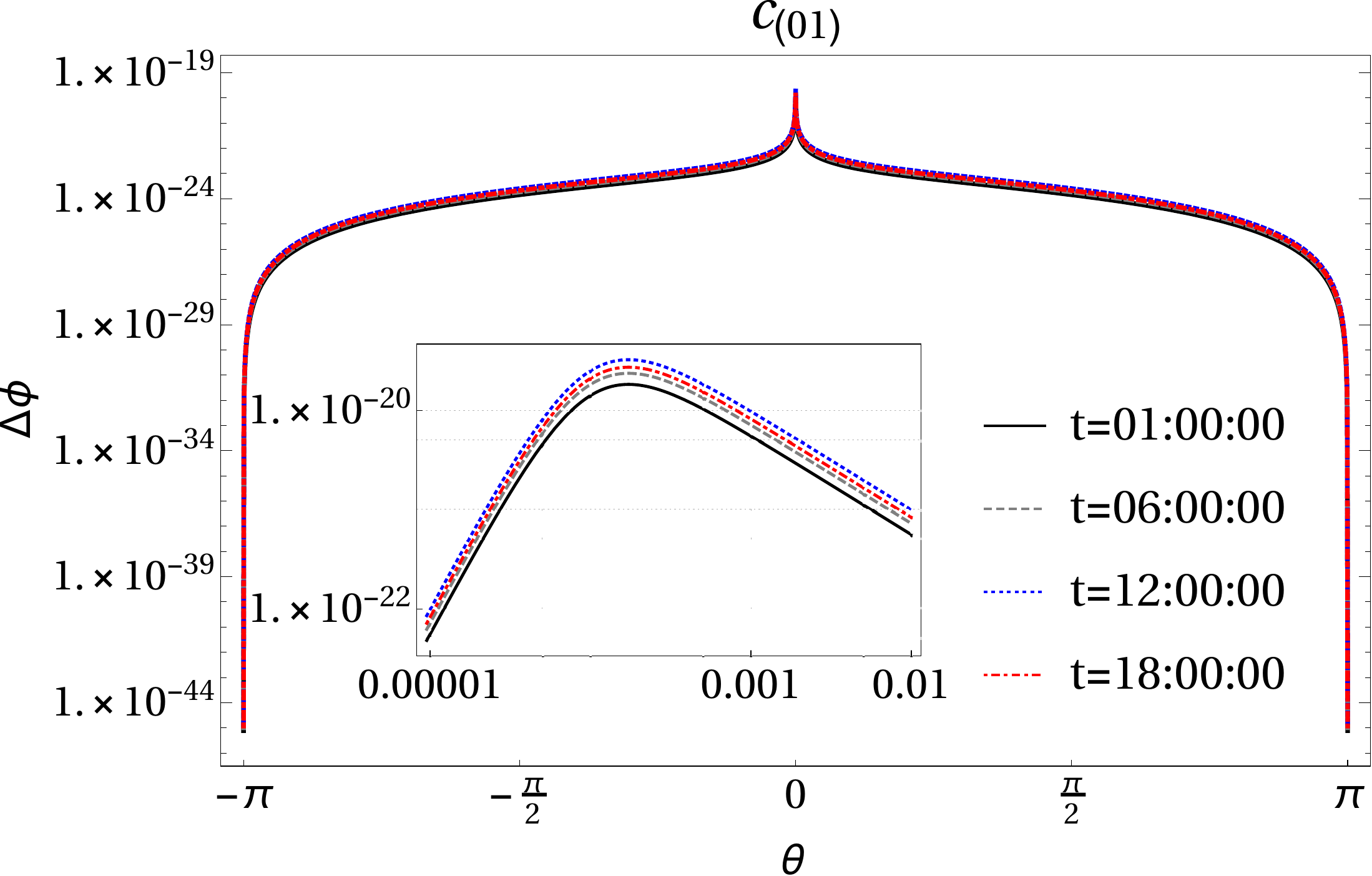}
    			\includegraphics[width=0.3297 \linewidth]{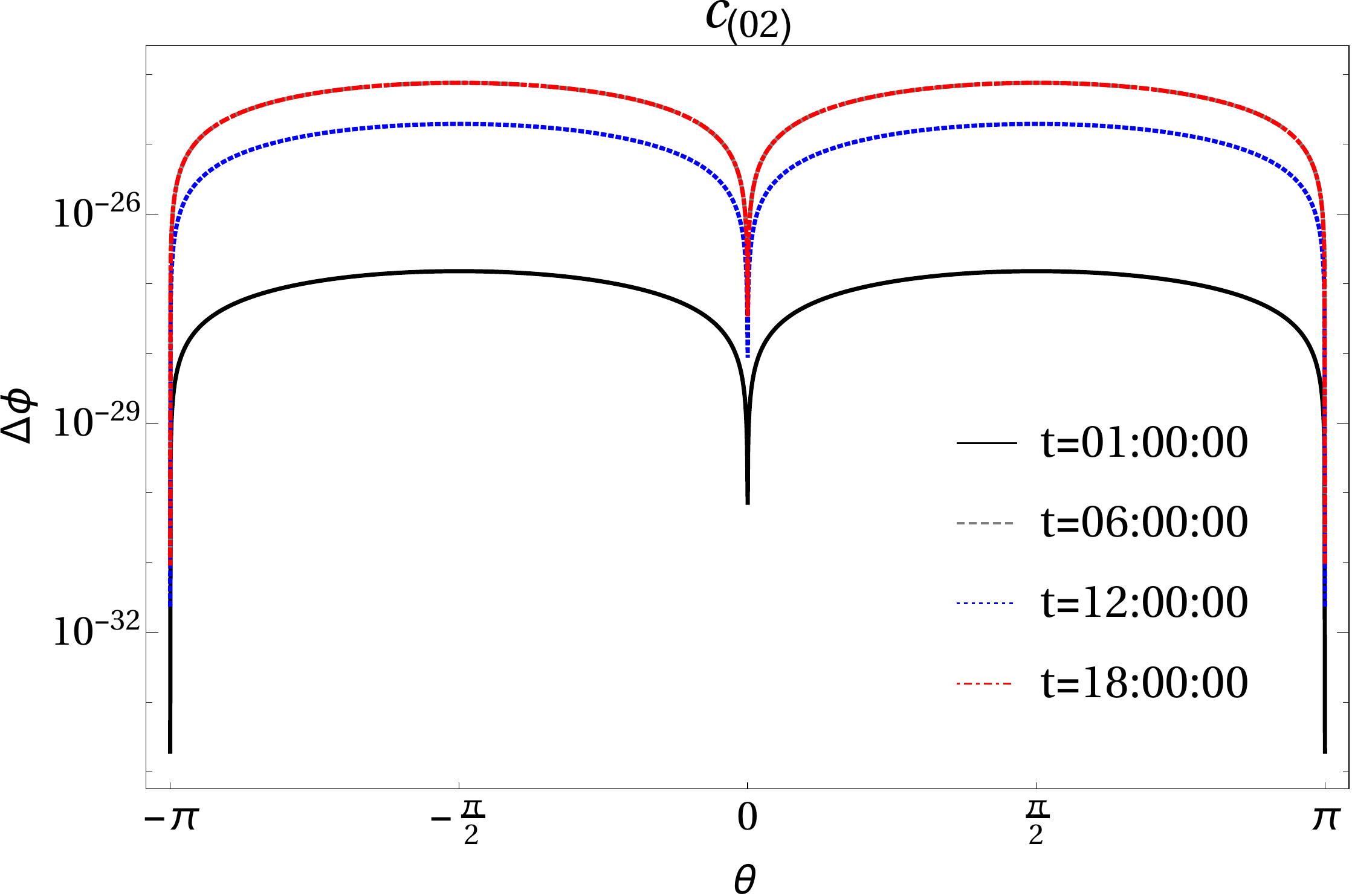}\\
    			\hspace{-1cm}
    			\includegraphics[width=0.3297 \linewidth]{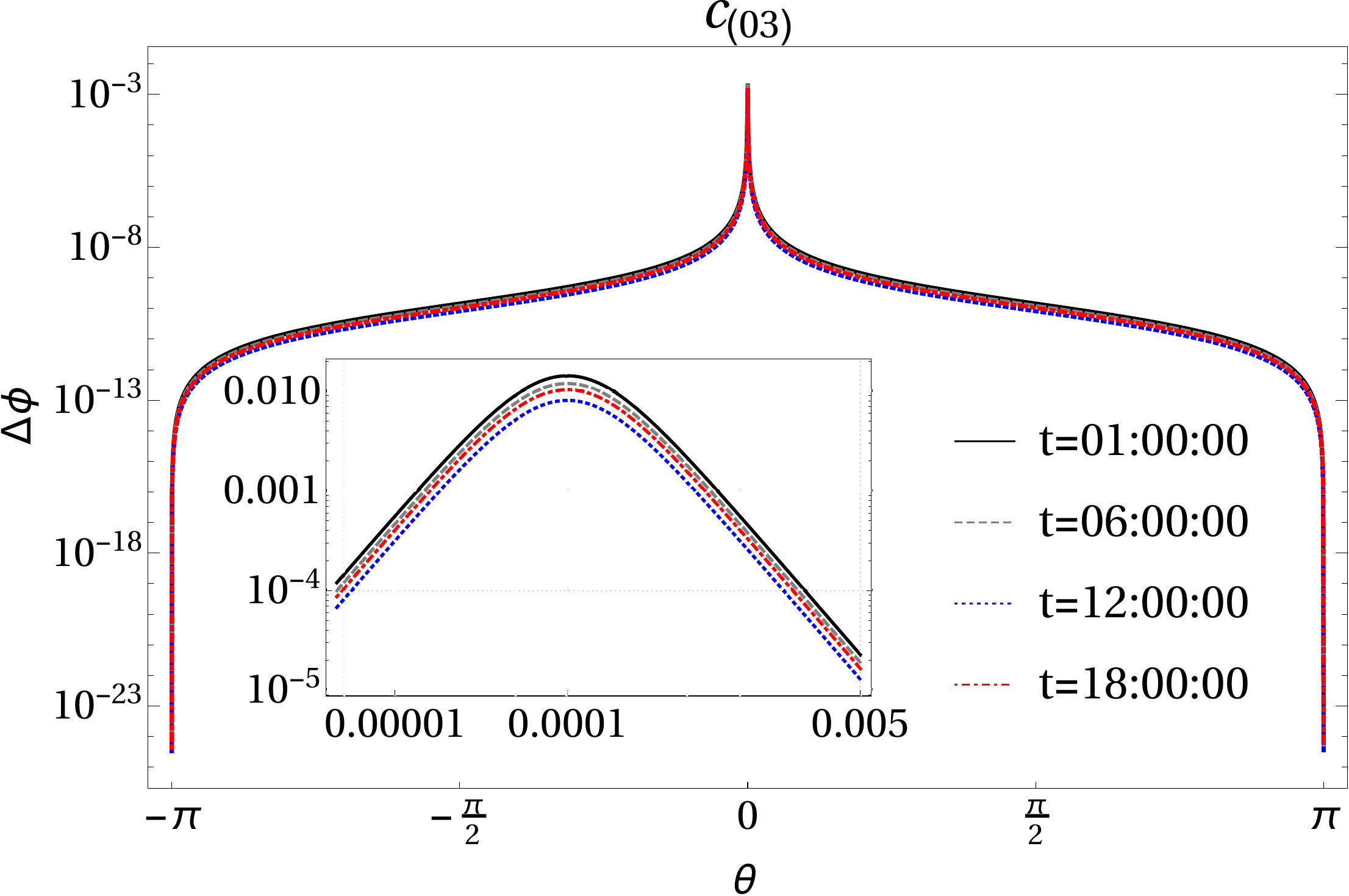}
    			\includegraphics[width=0.3297 \linewidth]{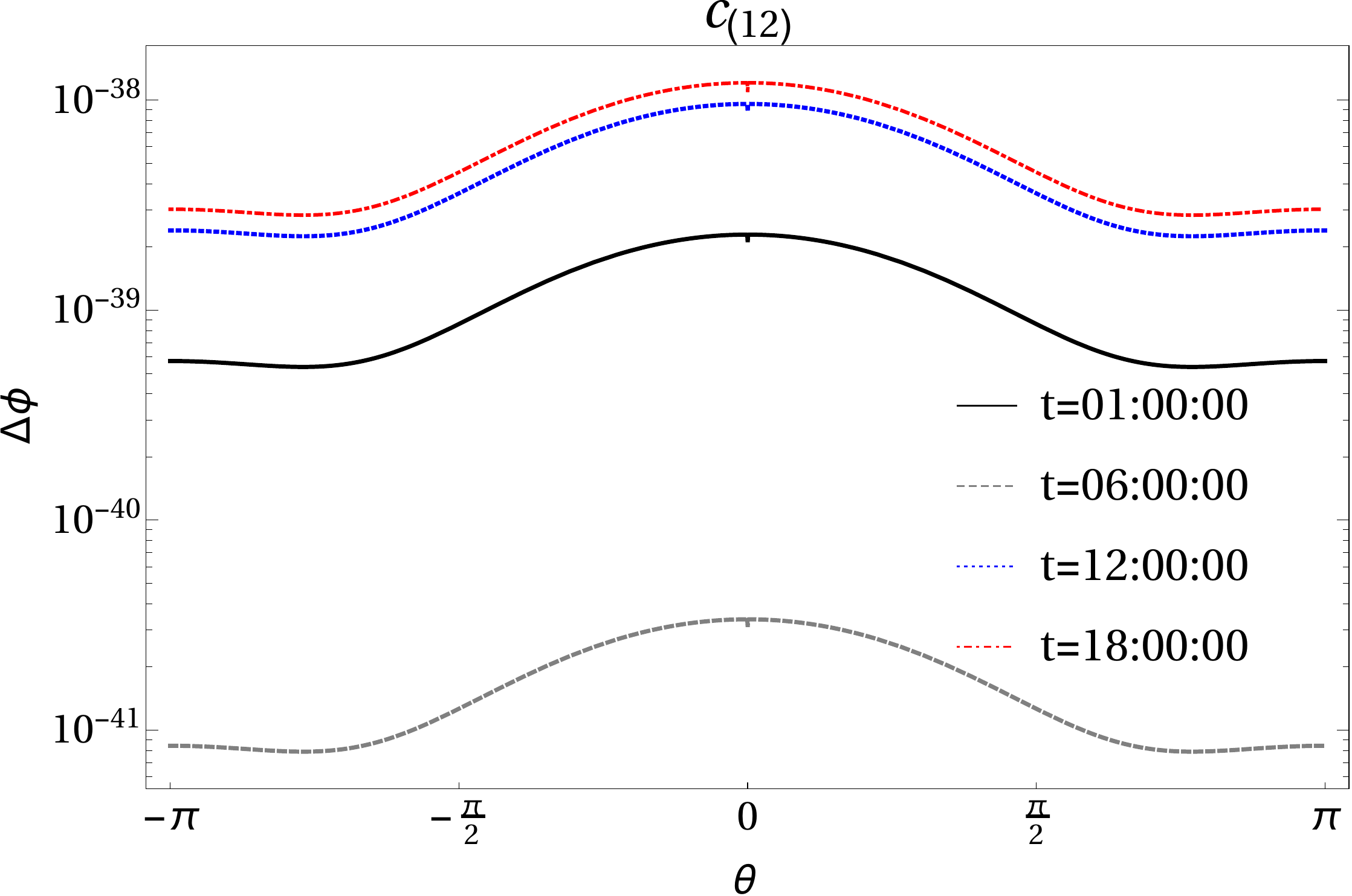}
    			\includegraphics[width=0.3297 \linewidth]{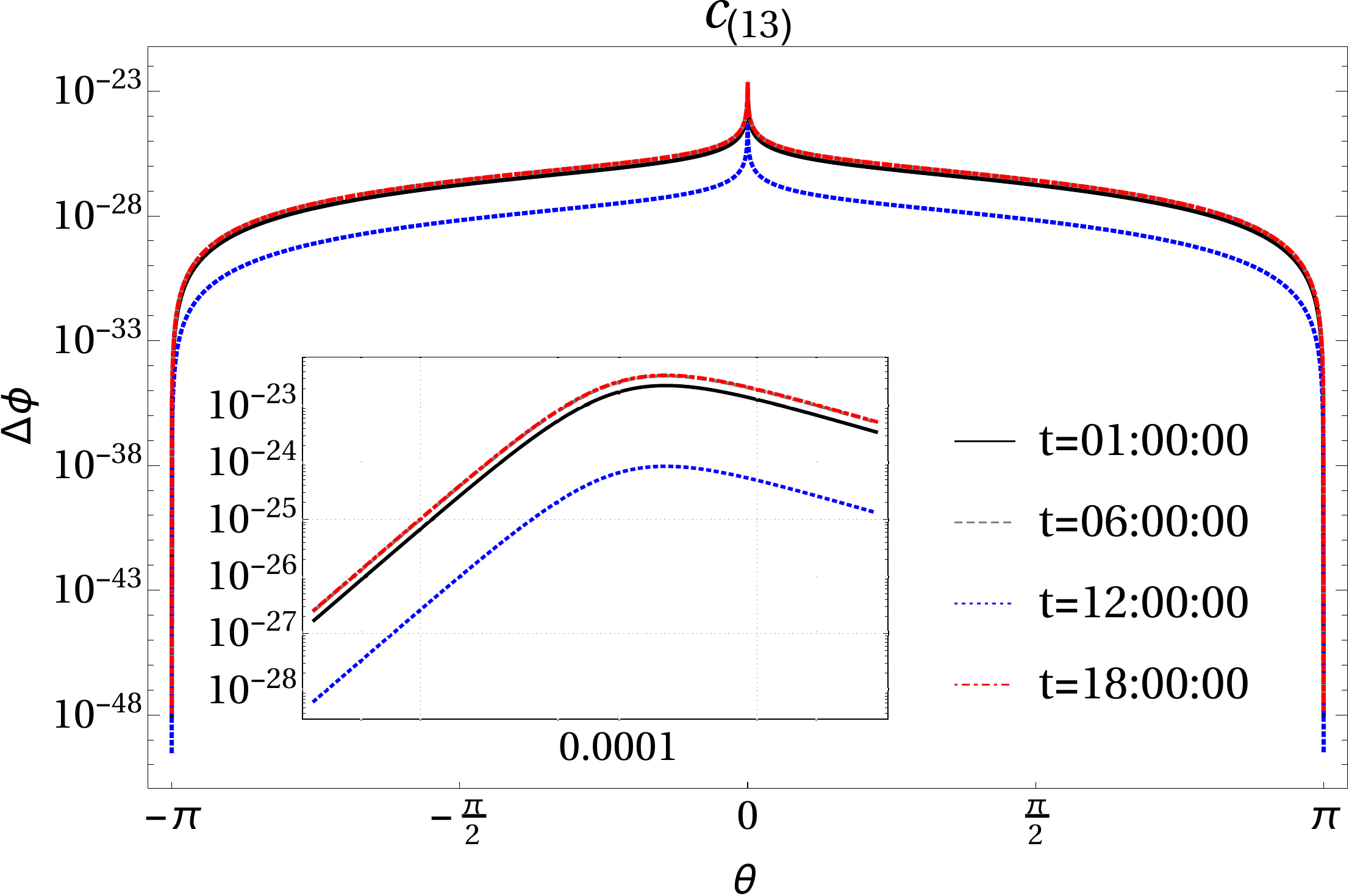}\\
    			\hspace{-1cm}
    			\includegraphics[width=0.3297 \linewidth]{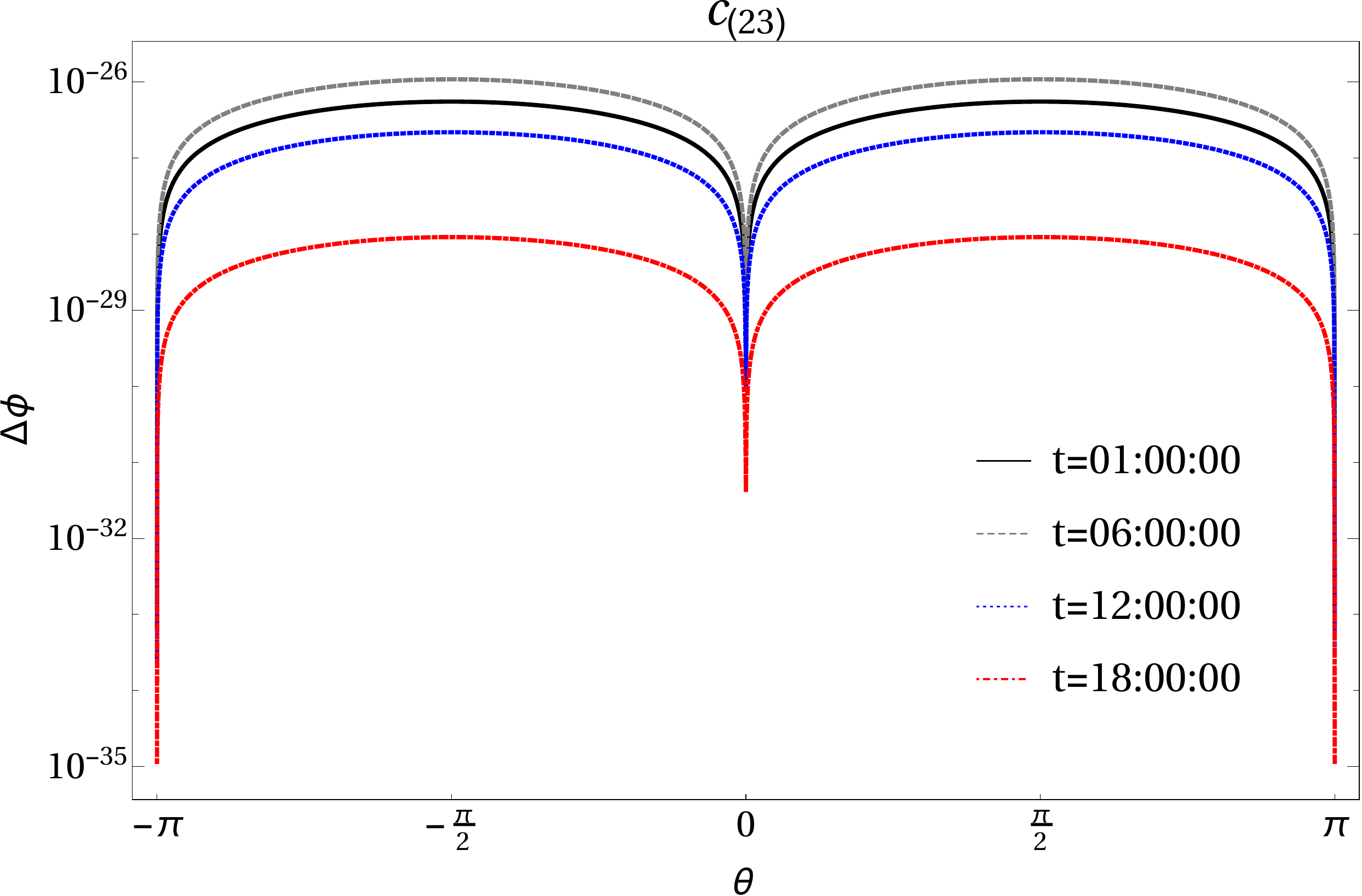}
    			\includegraphics[width=0.3297 \linewidth]{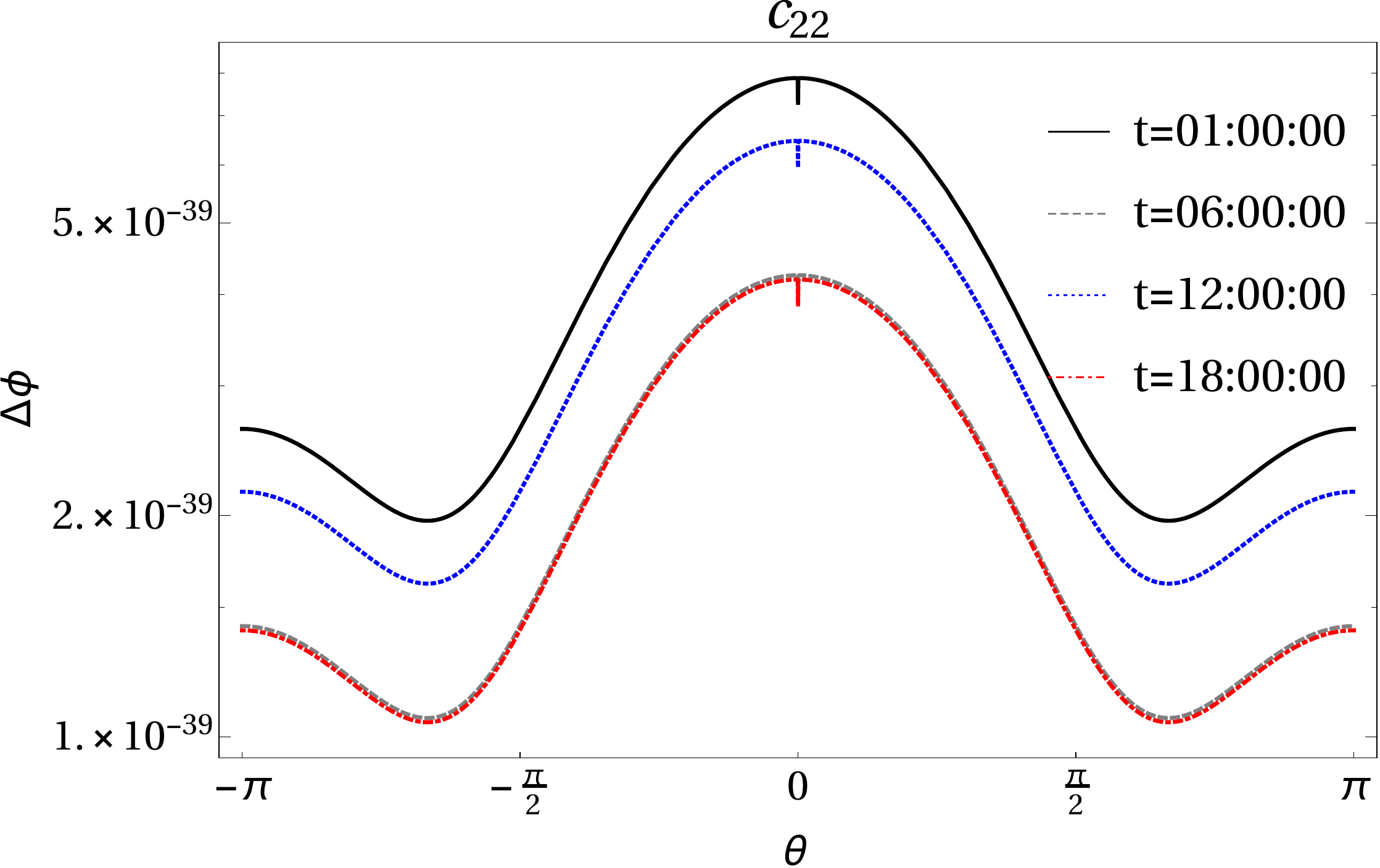}
    			\includegraphics[width=0.3297 \linewidth]{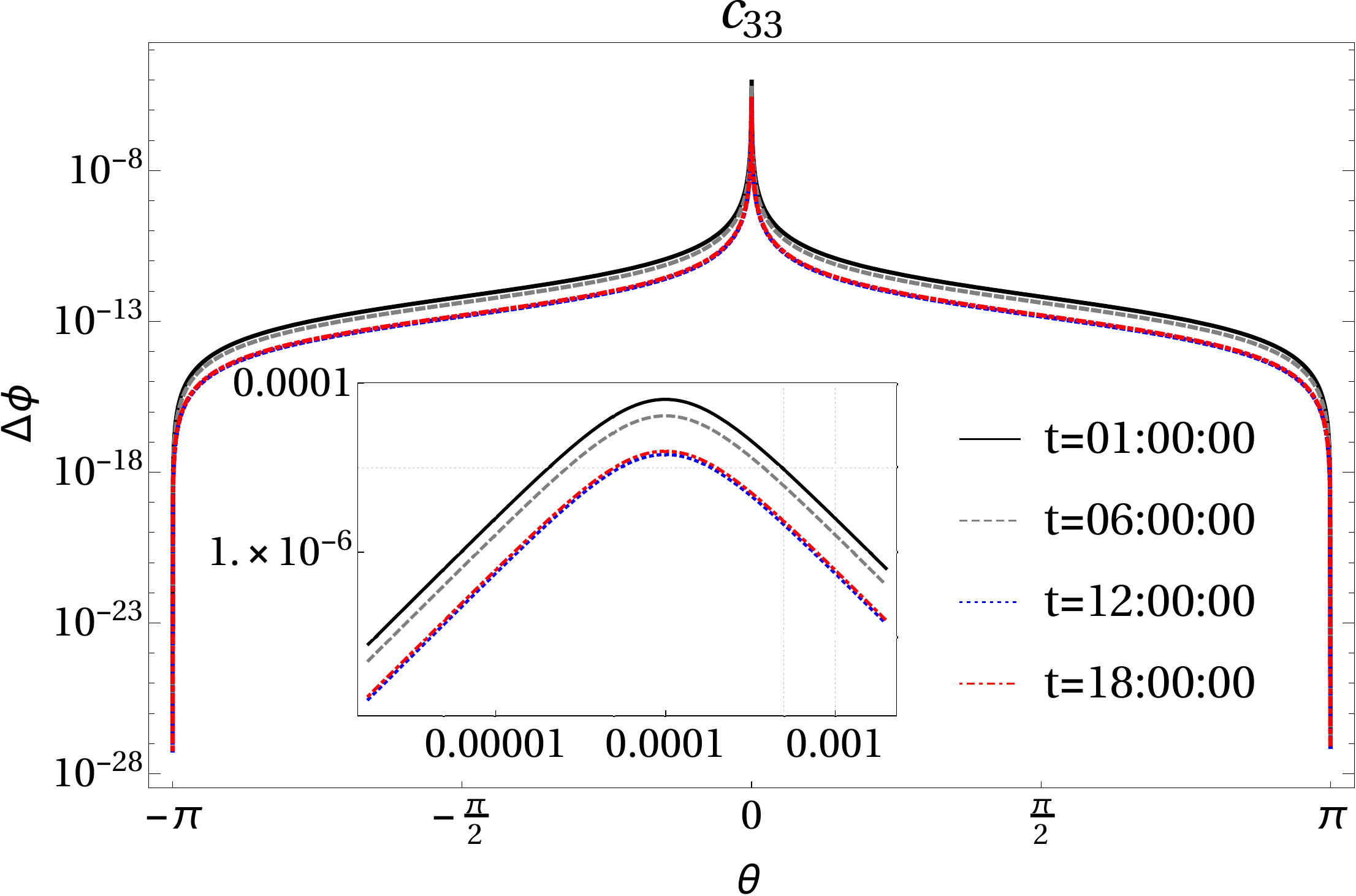}
    			\caption{\small FC phase, $\Delta \phi|_\mu^{\text{LV}}$ as a function of scattering angle $\theta$ for four different times in a day-night. We have taken $\varphi=\chi=\frac{\pi}{4}$.}
    			\label{fig:deltaphimu}
    		\end{figure}
    The time dependence of FC with the muon beam is more tangible compared to the electron beam. \\Moreover, we determine the FC phase for different components of $c_{\mu \nu}$ with the intention of exploring the effect of each component of $c_{\mu\nu}$ coefficients on FC phase.  To do that we assume that only one specific component of $c_{\mu\nu}$ contributes in the FC phase. Therefore, the FC phase $\Delta\phi|_e^{\text{LV}}$ for different components of $c_{\mu\nu}$  in forward scattering of the linearly polarized laser beam with the energy $ k_0\sim 0.1$ eV and electron beam as a function of scattering angle $\theta$ for four scenarios of assumptions on time free parameter: (time: 1,6,12,18) o'clock are plotted in Fig. \ref{fig:deltaphie}. Besides, Fig. \ref{fig:deltaphimu} represents the FC phase $\Delta\phi|_\mu^{\text{LV}}$ versus $\theta$ with the same assumptions for the laser beam interacting with the muon beam. For simplicity we set $\varphi=\chi=\pi/4$ and polarization of incoming laser beam as $U_0 = Q_0=I_0/2$ in  Figs. \ref{fig:deltaphie} and \ref{fig:deltaphimu}. $I_0$ is the intensity of the laser beam.
   	 
	%%%%%%%%%%%%%%%%%%%%%%%%%%%%%%%%%%%%%%%%%%%
	\section{Discussion and conclusion}\label{sec:V}
	%%%%%%%%%%%%%%%%%%%%%%%%%%%%%%%%%%%%%%%%%%%%%%
		We finally review how the laser beam interacting with the charged lepton beams can provide a new situation to constrain $c_{\mu \nu}$ coefficients. The $d_{\mu\nu}$ coefficient, as noted previously, has no contribution for the generation of circular polarization via forward scattering of laser and charged lepton beams.\par
As our results depend on the direction of the beams, the time of performing the experiment and location of the lab, there might be an optimal beam direction, performing time and position for the lab to observe the $c_{\mu \nu}$ effects.
\begin{table}[t]
	\centering
	\caption{\small {Location dependence of the independent components of $c_{\mu\nu}$. }}
	%\hspace{-2cm}
	\vspace{0.3cm}
	\scalebox{0.9}{
		\begin{tabular}{lll}
			\hline\hline
			Parameters&~~~~~~~Location Dependence\\
			\hline
			~~~$c_{00}$&~~~~~~~~~~$c_{TT}$\\
			%\hline
			%	~~~$c_{xx}$&~~~~~~~~~~$\frac{1}{2}\left((c_{XX}+c_{YY}) \cos ^2\chi +2 c_{ZZ} \sin ^2\chi\right)$~\\
			\hline
			~~~$c_{22}$&~~~~~~~~~~$\frac{c_{XX}+c_{YY}}{2}$\\
			\hline
			~~~$c_{33}$&~~~~~~~~~~$\frac{1}{2} \left((c_{XX}+c_{YY}) \sin ^2(\chi )+2 c_{ZZ} \cos ^2(\chi )\right)$\\
			\hline
			~~~$c_{(12)}$&~~~~~~~~~~$0$~\\
			\hline
			~~~$c_{(13)}$&~~~~~~~~~~$\sin\chi  \cos\chi  (c_{XX}+c_{YY}-2 c_{ZZ})$~\\
			\hline
			~~~$c_{(23)}$&~~~~~~~~~~$ 0$~\\
			\hline
			~~~$c_{(01)}$&~~~~~~~~~~$  - c_{(TZ)}\sin\chi $~\\
			\hline
			~~~$c_{(02)}$&~~~~~~~~~~$ 0$~\\
			\hline
			~~~$c_{(03)}$&~~~~~~~~~~$ c_{(TZ)} \cos\chi $~\\
			\hline\hline \\
		\end{tabular}
	} \label{tab:Conversion2}
\end{table}
As an example, the FC phases are obtained for four scenarios with assumptions on time with $\phi=\chi=\pi/4$. The energy of the linearly polarized laser beam is set $0.1$eV  and the energy of the charged lepton beam is assumed $q_0=1$ TeV, the number densities of the electron and the muon beam are $n_e=10^{10}$ cm$^{-3}$ and $n_\mu=10^{12}$cm$^{-3}$, respectively. Note these suggested charged beams are experimentally proposed \cite{{Delahaye:2013jla},{Barklow:2015tja},{CLIC:2016zwp}}.
 The results are given in Figs. \ref{fig:deltaphie} and \ref{fig:deltaphimu} for four different times during the day-night.
 Plots show that all FC phases $\Delta\phi|_{e,\mu}^{\text{LV}}$ get their maximum values at specific scattering angles for different components of $c_{\mu\nu}$. This could  be one  of the most important characters of our results to distinguish the contribution of the LV effect from other sources of circular polarization.\par
Based on the current constraint on the $c_{\mu\nu}$ components \cite{Kostelecky:2008ts} and the available sensitivity level to detect circular polarization or FC phase \cite{{Alexander:2008fp},{Smith:2016jqs},{Seto:2006hf}}, the estimated total FC phase is in the range of current experimental precision. It should be emphasized that  we do not need to use very high-intensity laser beams which would help us to avoid other background effects. \par
Moreover, as seen from Fig.\ref{fig:deltaphitot}, the magnitude of the FC phase is very slightly varying at different times of day-night. Besides, the experimental time-dependent data is not currently available. Therefore, for presenting attainable results with current technology,  averaging over the time seems reasonable. For reader's convenience the components of $c_{\mu\nu}$, after time averaging, in Earth coordinates and on the non-rotating frame, are given in Tab.\ref{tab:Conversion2}.  According to Tab.\ref{tab:Conversion2}, the coefficients $c_{00}$ and $c_{22}$ after time averaging no longer depend on Earth's latitude and the laboratory location $\chi$.  Time averaging of the $c_{(12)},c_{(23)},c_{(01)}$ components leads to vanishing values as  $<\cos\Omega t>$ and $<\sin\Omega t>$  are equal to zero.  However, as depicted in Figs. \ref{fig:deltaphie} and \ref{fig:deltaphimu},  the magnitudes of those components are very small before time averaging and out of reach with current sensitivity. \par 
	 We estimate the bounds on a combination of $c_{\mu \nu}$  components contributing to the FC phase based on available accuracy on  the FC phase $\Delta \phi \sim 10^{-3}-10^{-2}$rad \cite{MacDonald:2016iwi}. 
	 Assuming the same typical linearly polarized laser beam interacting with the electron beam as mentioned above and $\phi=\chi=\pi/4$, we find a bound of $4.35\times 10^{-15}$ on $[c_{TT}+1.4~c_{(TZ)}+0.25(c_{XX}+c_{YY}+2~c_{ZZ})]$. For the muon beam we obtain a looser bound of $3.9\times 10^{-13}$ on $[c_{TT}+1.4~c_{(TZ)}+0.25(c_{XX}+c_{YY}+2~c_{ZZ})]$. While the LV components in Eq. \ref{c-coefficients}  depend on both location of laboratory and time of the experiment, it would be possible to find various constraints on a combination of the $c_{\mu\nu}$ components at different locations and times. Moreover, by improving the sensitivity of the experiment to FC in the future, the obtained bounds will be improved.
	 \par
We should also mention that any backgrounds can be a possible source to generate circular polarization. For example, in addition to the LV correction to Compton scattering, interaction of photon with a LV background can also create circular polarization which is linearly proportional to the LV coefficient for the  photon sector of $k_F$ \cite{Bavarsad:2009hm} and $k_{AF}$ \cite{Alexander:2008fp}. This effect does not modify the Compton scattering but the dynamics of the laser beam can be influenced \cite{Bavarsad:2009hm}. Briefly we ensure that our study provides a valuable supplement to other theoretical and experimental frameworks for improving the available constraints on the LV coefficients.

	%%%%%%%%%%%%%%%%%%%%%%%%%%%%%%%%      Acknowledgments   %%%%%%%%%%%%%%%%%%%%%%%%%%%%%%%%%%%
	\section*{Acknowledgments}
	%%%%%%%%%%%%%%%%%%%%%%%%%%%%%%%%%%%%%%%%%%%%%%%%%%%%%%%%%%%%%%%%%%%%%%%%%%%%%%%%%%
	We are thankful to I.Motie for collaboration in the initial stages of this work. S.Tizchang is grateful to  S.Aghababaei for fruitful discussions on LV, and to  M.Aslaninejad for helpful discussion on charged lepton beams.
	\\

\end{document}